\documentstyle[12pt]{article}\newlength\Cscr\newlength\Csave\newlength
\Ctenthex\newlength\CFxsize\newlength
\CFxsizeps\newlength\CFsizemakebox\newlength\CFleftcrop\newlength
\CFrightcrop\newlength\CZtbldist\newlength\CZfigdist\newlength\CGDnum
\newlength\CGDtext\newcounter{Cscr}\newcounter
{CBcit}\newcounter{CFlette%
r}\newcounter{Ceqin%
dent}\newcounter{CBtnc}\newcounter{CBtntc}%
\newcounter{CEht}\newcounter{Cbscurr}%
\newcounter{CbsA}\newcounter{CbsB}\newcounter{CbsC}\newcounter{C%
bsD}
\hoffset\Cscr
\voffset\Cscr
\protect
\begin{document}\setcounter{figure}{1}\setcounter{table}%
{0}\renewcommand\theequation{\arabic{equation}}\renewcommand
\thetable{\arabic{table}}\renewcommand\thefigure{\arabic{figure}%
}\renewcommand\thesection{\arabic{section}}\renewcommand
\thesubsection{\arabic{section}.\arabic{subsection}}\renewcommand
\thesubsubsection{\arabic{section}.\arabic{subsection}.\arabic{s%
ubsubsection}}\setcounter{CEht}{10}\setcounter{CbsA}{1}%
\setcounter{CbsB}{1}\setcounter{CbsC}{1}\setcounter{CbsD}{1}{%
\centering{\protect\mbox{}}\\*[\baselineskip]{\large\bf The alia%
sing problem in lattice field theory}\\*}{\centering{\protect
\mbox{}}\\John P.~Costella\vspace{1ex}\\*}{\centering{\small\mbox
{}\protect\/{\protect\em Mentone Grammar, 63 Venice Street, Ment%
one, Victoria 3194, Australia\protect\/}}\\}{\centering{\small
\mbox{}\protect\/{\protect\em jpcostella@hotmail.com;\hspace{1ex%
} jpc@mentonegs.vic.edu.au;\hspace{1ex} www.ph.unimelb.edu.au/$%
\sim$jpc\protect\/}}\\}{\centering{\protect\mbox{}}\\(25\ April 
2004\vspace{1ex})\\}\par\vspace\baselineskip\setlength{\Csave}{%
\parskip}{\centering\small\bf Abstract\\}\setlength{\parskip}{-%
\baselineskip}\begin{quote}\setlength{\parskip}{0.5\baselineskip
}\small\noindent The intrinsically nonlinear nature of quantum f%
ield theory provides a fundamental complication for lattice calc%
ulations, when the physical implications of the subtleties of Fo%
urier theory are taken into account. Even though the fundamental 
fields are constrained to the first Brillouin zone, Fourier theo%
ry tells us that the high-momentum components of products of the%
se fields ``bleed into'' neighbouring Brillouin zones, where the%
y ``alias'' (or ``masquerade'') as low-momentum contributions, v%
iolating the conservation of energy and momentum, and fundamenta%
lly distorting calculations. In this paper I offer a general str%
ategy for eliminating the artefacts of aliasing in practical cal%
culations. \end{quote}\setlength{\parskip}{\Csave}\par
\refstepcounter{section}\vspace{1.5\baselineskip}\par{\centering
\bf\thesection. Introduction and motivation\\*[0.5\baselineskip]%
}\protect\indent\label{sect:Introduction}There is a strange rela%
tionship between the ways that engineers and theoretical physici%
sts employ Fourier theory. Engineers generally use it for physic%
al applications that are, from the theoretical physicist's point 
of view, almost trivally simple in structure: physical equations 
that are often linear, or, at worst, nonlinear in ways that are 
simple to understand, and even simpler to describe mathematicall%
y. Physicists, in contrast, analyse almost intractably complicat%
ed mathematical descriptions of physical reality, for which Four%
ier techniques are but one of the fundamental tools in what can 
be a vast mathematical toolbox of almost unbelievable sophistica%
tion and abstraction.\par It is ironic, then, that the average e%
ngineer often gets a more thorough grounding in the fundamentals 
of Fourier theory than the average theoretical physicist. I don'%
t know why this is so; it seems to be something of a historical 
accident. That the same mathematical formalism has been develope%
d, independently, in both fields is illustrated most starkly by 
the fact that the fundamental theorems and constructs of Fourier 
theory are named after different people, depending on which facu%
lty department is teaching it!\par It is perhaps a truism that e%
ngineers can spend much more time analysing every subtlety of th%
e physical theory they are using to describe their creations, si%
mply because such descriptions \mbox{}\protect\/{\protect\em are%
\protect\/} so fundamentally simple and straightforward (from th%
e physicist's point of view, anyway). Nevertheless, it seems to 
me that there are some lessons that engineers have long learnt, 
from their extensive application of Fourier theory to real-world 
applications (when a simple mathematical oversight can mean the 
difference between a device working as designed or failing disma%
lly), that have not been sufficiently hammered home to theoretic%
al physicists---if they have, indeed, been explicitly recognised 
as problems at all.\par I believe that the most insidious of the%
se is rooted in the simple process of forming products of fields 
in lattice field theory. It is a fundamental theorem of Fourier 
theory that forming the product of two fields in position space 
leads to a ``convolution'' of their momentum-space representatio%
ns. This process axiomatically leads to the Fourier transform of 
the product ``bleeding out'' of what the physicist calls the ``f%
irst Brillouin zone'' (the engineer calls it ``above the Nyquist 
frequency'') into the surrounding ``zones''\mbox{$\!$}, which ne%
cessarily means that it is automatically ``shifted down'' in mom%
entum. The engineer calls this phenomenon ``aliasing'': high-mom%
entum components ``masquerade'' as low-momentum components, and 
generally completely destroy the fidelity of the low-frequency s%
ignal in the process. The physicist generally describes the resu%
lt as being analogous to an ``Umklapp process'' in a crystal, be%
cause this is the physical example (with \mbox{}\protect\/{%
\protect\em real\protect\/} lattices, no less) for which this ph%
enomenon is most familiar.\par No matter what it is called, this 
phenomenon can be insidiously devastating for any lattice calcul%
ation aiming to obtain as accurate a result as possible for a gi%
ven amount of computing power. The fact that this phenomenon vio%
lates the conservation of energy and momentum should be sufficie%
nt to ring alarm bells for any theorist (there is no physically 
real ``crystal'' to supply the Umklapp momentum in quantum field 
theory---it just comes from nowhere!). That fields end up being 
in the wrong place in momentum space is of concern not just to t%
he theorist, but also the pragmatist who simply wants to extract 
physically meaningful numbers from a lattice calculation.\par A 
simple example will suffice to demonstrate the general havoc wre%
aked by aliasing. Imagine that we wish to compute some sort of e%
xpectation value, that depends on a product of some number of fu%
ndamental fields, together with a number of operators acting on 
the fields. Computing the expectation value corresponds to evalu%
ating the Fourier transform of this result at zero momentum. How%
ever, if aliasing is not prevented, we find that there are \mbox
{}\protect\/{\protect\em two\protect\/} contributions to the res%
ult. The first is the true zero-momentum result due to the latti%
cised approximation of the physical system (with whatever unavoi%
dable approximations and inaccuracies that that entails), which 
is what we are trying to extract. The second is due to the inter%
action of the components of each field (and the operators in que%
stion) \mbox{}\protect\/{\protect\em at high momentum values%
\protect\/}, which has been aliased down to zero momentum due to 
the fundamental properties of Fourier theory. The result, of cou%
rse, comes to us as a single number, with these two contribution%
s inextricably intertwined.\par It may come as a shock, to some, 
that this second contribution is present at all, and is to a gre%
ater or lesser extent confounding the entire lattice calculation 
by its presence. Indeed, the components that make up the second 
contribution rightly belong at \mbox{}\protect\/{\protect\em hig%
h\protect\/} momentum values---outside the first Brillouin zone 
completely. What are they doing back at zero momentum?\par Such 
is the fun and games that one encounters if one entrusts Fourier 
theory with our calculations, without first ``reading the fine p%
rint'' of that mathematical formalism. In the rest of this paper%
, I will try to provide a ``user's guide'' for Fourier theory, i%
n which this ``fine print'' is enlarged, and made intelligible. 
 For many readers, much of what I will review here is like ``kin%
dergarten physics''---not worried about since undergraduate days%
; but when it is all put together, the importance of its ramific%
ations for lattice calculations may be surprising. Other readers 
may find useful the collecting together of the fundamental theor%
ems of Fourier theory, and its connection with physical reality, 
which is often developed somewhat flippantly and ``on the run'' 
in elementary physics applications. Finally, I will offer a gene%
ral strategy for removing the aliasing problem at its root, that 
may be of use in practical lattice calculations.\par
\refstepcounter{section}\vspace{1.5\baselineskip}\par{\centering
\bf\thesection. A review of the fundamentals of Fourier theory%
\\*[0.5\baselineskip]}\protect\indent\label{sect:Fourier}In this 
section I provide a brief overview of the fundamental mathematic%
al elements of Fourier theory. I concentrate on one-dimensional 
lattices; the extension to an arbitrary number of dimensions is 
elementary, and does not introduce any new constructs. The physi%
cal application and interpretation of this purely mathematical m%
achinery will be discussed in the next section.\par We start wit%
h a set of $N$ lattice sites. We can label these sites with an i%
ntegral index $n$ that falls within a range of $N$ successive in%
tegers. For example, we might have $n$ running from $0$ to 
\raisebox{0ex}[1ex][0ex]{$\protect\displaystyle N\:\!\!-1$}; but 
any other such range is equally possible. For definiteness, for 
the rest of this section we will assume that $n$ runs from $1$ t%
o $N$.\par We can define a function, say, \raisebox{0ex}[1ex][0e%
x]{$\protect\displaystyle f_{\:\!\!n}$}, that is defined on each 
of the $N$ lattice sites. The $N$ values \raisebox{0ex}[1ex][0ex%
]{$\protect\displaystyle f_{\:\!\!n}$} are not restricted in any 
way at all. They may follow some mathematical rule---and in prac%
tical applications usually will---but in principle they may be s%
imply $N$ numbers created at random.\par We now define the \mbox
{}\protect\/{\protect\em Fourier transform\protect\/}, \raisebox
{0ex}[1ex][0ex]{$\protect\displaystyle F_b\equiv{{\cal F}}\{f_{%
\:\!\!n}\}$}, as follows: \setcounter{Ceqindent}{0}\protect\begin
{eqnarray}\protect\left.\protect\begin{array}{rcl}\protect
\displaystyle\hspace{-1.3ex}&\protect\displaystyle F_b\equiv\mbox
{$\protect\displaystyle\protect\frac{1}{\sqrt N}$}\sum_{n=1}^Ne^%
{-2\pi ibn/N\!}f_{\:\!\!n}.\setlength{\Cscr}{\value{CEht}%
\Ctenthex}\addtolength{\Cscr}{-1.0ex}\protect\raisebox{0ex}[%
\value{CEht}\Ctenthex][\Cscr]{}\protect\end{array}\protect\right
.\protect\label{eq:Fourier-Fourier}\protect\end{eqnarray}%
\setcounter{CEht}{10}The values $F_b$ represent lattice sites in 
``Fourier space''\mbox{$\!$}. The index $b$, like $n$, can fall 
within any range of $N$ successive integers, which need not matc%
h the range of $n$. For definiteness, however, for the rest of t%
his section we will assume that $b$ also runs from $1$ to $N$.%
\par The fundamental theorem of Fourier theory is that the funct%
ion \raisebox{0ex}[1ex][0ex]{$\protect\displaystyle f_{\:\!\!n}$%
} can be retrieved from the function $F_b$ by means of the \mbox
{}\protect\/{\protect\em inverse Fourier transform\protect\/}, 
\raisebox{0ex}[1ex][0ex]{$\protect\displaystyle f_{\:\!\!n}\equiv
{{\cal F}}^{-1\:\!\!}\{F_b\}$}: \setcounter{Ceqindent}{0}\protect
\begin{eqnarray}\protect\left.\protect\begin{array}{rcl}\protect
\displaystyle\hspace{-1.3ex}&\protect\displaystyle f_{\:\!\!n}%
\equiv\mbox{$\protect\displaystyle\protect\frac{1}{\sqrt N}$}\sum
_{b=1}^Ne^{+2\pi ibn/N}F_b,\par\setlength{\Cscr}{\value{CEht}%
\Ctenthex}\addtolength{\Cscr}{-1.0ex}\protect\raisebox{0ex}[%
\value{CEht}\Ctenthex][\Cscr]{}\protect\end{array}\protect\right
.\protect\label{eq:Fourier-InverseFourier}\protect\end{eqnarray}%
\setcounter{CEht}{10}which differs from (\protect\ref{eq:Fourier%
-Fourier}) only in that $n$ and $b$ have been interchanged, $f$ 
and $F$ have been interchanged, and $-i$ has been replaced by $+%
i$ in the complex exponential. It is instructive to review the p%
roof of the fundamental theorem, which is straightforward. We do 
this by evaluating the right-hand side of (\protect\ref{eq:Fouri%
er-InverseFourier}). Substituting $F_b$ from (\protect\ref{eq:Fo%
urier-Fourier}), we have \setcounter{Ceqindent}{0}\protect\begin
{eqnarray}\mbox{$\protect\displaystyle\protect\frac{1}{\sqrt N}$%
}\sum_{b=1}^Ne^{+2\pi ibn/N}F_b\hspace{-1.3ex}&\displaystyle=&%
\hspace{-1.3ex}\mbox{$\protect\displaystyle\protect\frac{1}{\sqrt
N}$}\sum_{b=1}^Ne^{+2\pi ibn/N}\mbox{$\protect\displaystyle
\protect\frac{1}{\sqrt N}$}\sum_{n^{\prime\!}=1}^Ne^{-2\pi ibn^{%
\prime\!}/N}f_{\:\!\!n'}\protect\nonumber\setlength{\Cscr}{\value
{CEht}\Ctenthex}\addtolength{\Cscr}{-1.0ex}\protect\raisebox{0ex%
}[\value{CEht}\Ctenthex][\Cscr]{}\\*[0ex]\protect\displaystyle
\hspace{-1.3ex}&\displaystyle=&\hspace{-1.3ex}\mbox{$\protect
\displaystyle\protect\frac{1}{N}$}\sum_{b=1}^N\sum_{n^{\prime\!}%
=1}^Ne^{2\pi ib(n-n')/N}f_{\:\!\!n'}\protect\nonumber\setlength{%
\Cscr}{\value{CEht}\Ctenthex}\addtolength{\Cscr}{-1.0ex}\protect
\raisebox{0ex}[\value{CEht}\Ctenthex][\Cscr]{}\\*[0ex]\protect
\displaystyle\hspace{-1.3ex}&\displaystyle=&\hspace{-1.3ex}\mbox
{$\protect\displaystyle\protect\frac{1}{N}$}\sum_{n^{\prime\!}=1%
}^Nf_{\:\!\!n'\!}\sum_{b=1}^Ne^{2\pi ib(n-n')/N\!},\protect\label
{eq:Fourier-LastLineProof}\setlength{\Cscr}{\value{CEht}\Ctenthex
}\addtolength{\Cscr}{-1.0ex}\protect\raisebox{0ex}[\value{CEht}%
\Ctenthex][\Cscr]{}\protect\end{eqnarray}\setcounter{CEht}{10}wh%
ere in the last line we have simply interchanged the order of pe%
rforming the sums, which is permissible because the sums are man%
ifestly finite. We now consider the last sum in (\protect\ref{eq%
:Fourier-LastLineProof}), \setcounter{Ceqindent}{0}\protect\begin
{eqnarray}\hspace{-1.3ex}&\displaystyle\sum_{b=1}^Ne^{2\pi ib(n-%
n')/N\!}.\protect\nonumber\setlength{\Cscr}{\value{CEht}\Ctenthex
}\addtolength{\Cscr}{-1.0ex}\protect\raisebox{0ex}[\value{CEht}%
\Ctenthex][\Cscr]{}\protect\end{eqnarray}\setcounter{CEht}{10}Th%
ere are two possible cases. If \raisebox{0ex}[1ex][0ex]{$\protect
\displaystyle n=n^{\prime\!}$}, the exponent will vanish for any 
value of $b$, and hence the complex exponential is unity for all 
values of $b$, and so we are simply summing up $N$ copies of uni%
ty, which is just $N$. If \raisebox{0ex}[1ex][0ex]{$\protect
\displaystyle n\neq n^{\prime\!}$}, on the other hand, it is str%
aightforward to verify that the sum vanishes by symmetry. In the 
case that \raisebox{0ex}[1ex][0ex]{$\protect\displaystyle|n-n'|$%
} is coprime to $N$, then we are simply adding together all the 
$N$-th roots of unity, which are equally spaced around the unit 
circle in the Argand plane, and hence sum (vectorially) to zero. 
In the alternative case that \raisebox{0ex}[1ex][0ex]{$\protect
\displaystyle|n-n'|$} is not coprime to $N$, then if \raisebox{0%
ex}[1ex][0ex]{$\protect\displaystyle g=\gcd(|n-n'|,N)$}, we are 
simply adding together all the $(N/g)$-th roots of unity, a tota%
l of $g$ times. Thus \setcounter{Ceqindent}{0}\protect\begin{eqn%
array}\protect\left.\protect\begin{array}{rcl}\protect
\displaystyle\hspace{-1.3ex}&\protect\displaystyle\sum_{b=1}^Ne^%
{2\pi ib(n-n')/N\!}=N^{\:\!}\delta_{nn'},\setlength{\Cscr}{\value
{CEht}\Ctenthex}\addtolength{\Cscr}{-1.0ex}\protect\raisebox{0ex%
}[\value{CEht}\Ctenthex][\Cscr]{}\protect\end{array}\protect
\right.\protect\label{eq:Fourier-KroneckerIdentity}\protect\end{%
eqnarray}\setcounter{CEht}{10}where \raisebox{0ex}[1ex][0ex]{$%
\protect\displaystyle\delta_{ij}$} is the Kronecker delta symbol%
. Substituting (\protect\ref{eq:Fourier-KroneckerIdentity}) into 
(\protect\ref{eq:Fourier-LastLineProof}) yields \setcounter{Ceqi%
ndent}{0}\protect\begin{eqnarray}\hspace{-1.3ex}&\displaystyle
\mbox{$\protect\displaystyle\protect\frac{1}{N}$}\sum_{n^{\prime
\!}=1}^Nf_{\:\!\!n'}{}^{\:\!}N^{\:\!}\delta_{nn'}\equiv f_{\:\!%
\!n},\protect\nonumber\setlength{\Cscr}{\value{CEht}\Ctenthex}%
\addtolength{\Cscr}{-1.0ex}\protect\raisebox{0ex}[\value{CEht}%
\Ctenthex][\Cscr]{}\protect\end{eqnarray}\setcounter{CEht}{10}he%
nce establishing the fundamental theorem. The theorem equally pr%
ovides that the $F_b$ may be uniquely recovered from the 
\raisebox{0ex}[1ex][0ex]{$\protect\displaystyle f_{\:\!\!n}$}. I%
ndeed, there is no fundamental mathematical distinction between 
the ``original'' space and ``Fourier'' space; they simply provid%
e two distinct but equivalent representations of $N$ complex qua%
ntities.\par The expressions for the Fourier transformation (%
\protect\ref{eq:Fourier-Fourier}) and its inverse (\protect\ref{%
eq:Fourier-InverseFourier}) show why the indices $n$ and $b$ may 
be shifted by any arbitrary integral multiple of $N$ without cha%
nging the mathematical results. Namely, if we replace $n$ by 
\raisebox{0ex}[1ex][0ex]{$\protect\displaystyle n+cN$}, where $c%
$ is an integer, then the complex exponential factor in (\protect
\ref{eq:Fourier-InverseFourier}) or (\protect\ref{eq:Fourier-Fou%
rier}) becomes \setcounter{Ceqindent}{0}\protect\begin{eqnarray}%
\hspace{-1.3ex}&\displaystyle e^{\pm2\pi ib(n+cN)/N}\equiv e^{\pm
2\pi ibn/N}e^{\pm2\pi ibc}\equiv e^{\pm2\pi ibn/N\!},\protect
\nonumber\setlength{\Cscr}{\value{CEht}\Ctenthex}\addtolength{%
\Cscr}{-1.0ex}\protect\raisebox{0ex}[\value{CEht}\Ctenthex][\Cscr
]{}\protect\end{eqnarray}\setcounter{CEht}{10}because \raisebox{%
0ex}[1ex][0ex]{$\protect\displaystyle e^{\pm2\pi ibc}\equiv1$} i%
f $b$ and $c$ are integers, as they are here. Likewise, if we re%
place $b$ by \raisebox{0ex}[1ex][0ex]{$\protect\displaystyle b+c%
N$}, then the complex exponential factor in (\protect\ref{eq:Fou%
rier-InverseFourier}) or (\protect\ref{eq:Fourier-Fourier}) beco%
mes \setcounter{Ceqindent}{0}\protect\begin{eqnarray}\hspace{-1.%
3ex}&\displaystyle e^{\pm2\pi i(b+cN)n/N}\equiv e^{\pm2\pi ibn/N%
}e^{\pm2\pi icn}\equiv e^{\pm2\pi ibn/N\!}.\protect\nonumber
\setlength{\Cscr}{\value{CEht}\Ctenthex}\addtolength{\Cscr}{-1.0%
ex}\protect\raisebox{0ex}[\value{CEht}\Ctenthex][\Cscr]{}\protect
\end{eqnarray}\setcounter{CEht}{10}\par Let us now consider what 
happens if we form the product of two functions \raisebox{0ex}[1%
ex][0ex]{$\protect\displaystyle f_{\:\!\!n}$} and $g_n$, 
\setcounter{Ceqindent}{0}\protect\begin{eqnarray}\hspace{-1.3ex}%
&\displaystyle(f\!g)_n\equiv f_{\:\!\!n}g_n.\protect\nonumber
\setlength{\Cscr}{\value{CEht}\Ctenthex}\addtolength{\Cscr}{-1.0%
ex}\protect\raisebox{0ex}[\value{CEht}\Ctenthex][\Cscr]{}\protect
\end{eqnarray}\setcounter{CEht}{10}If we take the Fourier transf%
orm of \raisebox{0ex}[1ex][0ex]{$\protect\displaystyle(f\!g)_n$}%
, we have, from (\protect\ref{eq:Fourier-Fourier}), \setcounter{%
Ceqindent}{0}\protect\begin{eqnarray}\hspace{-1.3ex}&%
\displaystyle{{\cal F}}\{(f\!g)_n\}=\mbox{$\protect\displaystyle
\protect\frac{1}{\sqrt N}$}\sum_{n=1}^Ne^{-2\pi ibn/N}f_{\:\!\!n%
}g_n.\protect\nonumber\setlength{\Cscr}{\value{CEht}\Ctenthex}%
\addtolength{\Cscr}{-1.0ex}\protect\raisebox{0ex}[\value{CEht}%
\Ctenthex][\Cscr]{}\protect\end{eqnarray}\setcounter{CEht}{10}If 
we now use (\protect\ref{eq:Fourier-InverseFourier}) to write 
\raisebox{0ex}[1ex][0ex]{$\protect\displaystyle f_{\:\!\!n}$} an%
d $g_n$ in terms of their Fourier transforms $F_b$ and $G_b$, th%
is becomes \setcounter{Ceqindent}{0}\protect\begin{eqnarray}%
\hspace{-1.3ex}&&\hspace{-1.3ex}\mbox{$\protect\displaystyle
\protect\frac{1}{\sqrt N}$}\sum_{n=1}^Ne^{-2\pi ibn/N}\mbox{$%
\protect\displaystyle\protect\frac{1}{\sqrt N}$}\sum_{b^{\prime
\!}=1}^Ne^{+2\pi ib^{\prime\!}n/N}F_{b^{{\prime\!}\,}}\mbox{$%
\protect\displaystyle\protect\frac{1}{\sqrt N}$}\sum_{b^{\prime
\:\!\!\prime\!}=1}^Ne^{+2\pi ib^{\prime\:\!\!\prime\!}n/N}G_{b^{%
{\prime\:\!\!\prime\!}\,}}\protect\nonumber\setlength{\Cscr}{%
\value{CEht}\Ctenthex}\addtolength{\Cscr}{-1.0ex}\protect
\raisebox{0ex}[\value{CEht}\Ctenthex][\Cscr]{}\\*[0ex]\protect
\displaystyle\hspace{-1.3ex}&\displaystyle&\hspace{-1.3ex}{%
\protect\mbox{}}\hspace{\value{Ceqindent}\Ctenthex}=\mbox{$%
\protect\displaystyle\protect\frac{1}{N^{3/2}}$}\sum_{n=1}^N\sum
_{b^{\prime\!}=1}^N\sum_{b^{\prime\:\!\!\prime\!}=1}^Ne^{2\pi i(%
b^{\prime\!}+b^{\prime\:\!\!\prime\!}-b)n/N}F_{b^{{\prime\!}\,}}%
G_{b^{{\prime\:\!\!\prime\!}\,}}\protect\nonumber\setlength{\Cscr
}{\value{CEht}\Ctenthex}\addtolength{\Cscr}{-1.0ex}\protect
\raisebox{0ex}[\value{CEht}\Ctenthex][\Cscr]{}\\*[0ex]\protect
\displaystyle\hspace{-1.3ex}&\displaystyle&\hspace{-1.3ex}{%
\protect\mbox{}}\hspace{\value{Ceqindent}\Ctenthex}=\mbox{$%
\protect\displaystyle\protect\frac{1}{N^{3/2}}$}\sum_{b^{\prime
\!}=1}^NF_{b^{\prime\!}}\sum_{b^{\prime\:\!\!\prime\!}=1}^NG_{b^%
{\prime\:\!\!\prime\!}}\sum_{n=1}^Ne^{2\pi i(b^{\prime\!}+b^{%
\prime\:\!\!\prime\!}-b)n/N\!},\protect\label{eq:Fourier-FinalLi%
neConvolution}\setlength{\Cscr}{\value{CEht}\Ctenthex}%
\addtolength{\Cscr}{-1.0ex}\protect\raisebox{0ex}[\value{CEht}%
\Ctenthex][\Cscr]{}\protect\end{eqnarray}\setcounter{CEht}{10}wh%
ere again we can interchange the order of the sums because they 
are all manifestly finite. We can now use the result (\protect
\ref{eq:Fourier-KroneckerIdentity}) for the final sum in (%
\protect\ref{eq:Fourier-FinalLineConvolution}), to obtain 
\setcounter{Ceqindent}{0}\protect\begin{eqnarray}\hspace{-1.3ex}%
&\displaystyle\mbox{$\protect\displaystyle\protect\frac{1}{N^{3/%
2}}$}\sum_{b^{\prime\!}=1}^NF_{b^{\prime\!}}\sum_{b^{\prime\:\!%
\!\prime\!}=1}^NG_{b^{\prime\:\!\!\prime\!}}{}^{\,}N^{\:\!}\delta
_{b^{{\prime\:\!\!\prime\!}\!},\,b-b^{\prime\!}}=\mbox{$\protect
\displaystyle\protect\frac{1}{\sqrt N}$}\sum_{b^{\prime\!}=1}^NF%
_{b'}G_{b-b^{\prime\!}}.\protect\nonumber\setlength{\Cscr}{\value
{CEht}\Ctenthex}\addtolength{\Cscr}{-1.0ex}\protect\raisebox{0ex%
}[\value{CEht}\Ctenthex][\Cscr]{}\protect\end{eqnarray}%
\setcounter{CEht}{10}This last bilinear operation in $F_b$ and $%
G_b$ is of such fundamental importance that it is given a specia%
l name, the \mbox{}\protect\/{\protect\em convolution\protect\/} 
of $F$ and $G$, and is denoted by a special symbol, the asterisk%
: \setcounter{Ceqindent}{0}\protect\begin{eqnarray}\protect\left
.\protect\begin{array}{rcl}\protect\displaystyle\hspace{-1.3ex}&%
\protect\displaystyle(F^{\:\!\!}\!\ast\!G)_b\equiv\mbox{$\protect
\displaystyle\protect\frac{1}{\sqrt N}$}\sum_{b^{\prime\!}=1}^NF%
_{b'}G_{b-b^{\prime\!}}.\setlength{\Cscr}{\value{CEht}\Ctenthex}%
\addtolength{\Cscr}{-1.0ex}\protect\raisebox{0ex}[\value{CEht}%
\Ctenthex][\Cscr]{}\protect\end{array}\protect\right.\protect
\label{eq:Fourier-ConvolutionFourier}\protect\end{eqnarray}%
\setcounter{CEht}{10}We have thus shown that the Fourier transfo%
rm of a product of functions is the \mbox{}\protect\/{\protect\em
convolution\protect\/} of their Fourier transforms: \setcounter{%
Ceqindent}{0}\protect\begin{eqnarray}\protect\left.\protect\begin
{array}{rcl}\protect\displaystyle\hspace{-1.3ex}&\protect
\displaystyle{{\cal F}}\{(f\!g)_n\}=(F^{\:\!\!}\!\ast\!G)_b.%
\setlength{\Cscr}{\value{CEht}\Ctenthex}\addtolength{\Cscr}{-1.0%
ex}\protect\raisebox{0ex}[\value{CEht}\Ctenthex][\Cscr]{}\protect
\end{array}\protect\right.\protect\label{eq:Fourier-ProductConvo%
lution}\protect\end{eqnarray}\setcounter{CEht}{10}It is straight%
forward to verify that the same theorem holds true for the inver%
se Fourier transform of a product of functions in Fourier space: 
\setcounter{Ceqindent}{0}\protect\begin{eqnarray}\protect\left.%
\protect\begin{array}{rcl}\protect\displaystyle\hspace{-1.3ex}&%
\protect\displaystyle{{\cal F}}^{-1\:\!\!}\{(FG)_b\}=(f\!\ast\!g%
)_n,\setlength{\Cscr}{\value{CEht}\Ctenthex}\addtolength{\Cscr}{%
-1.0ex}\protect\raisebox{0ex}[\value{CEht}\Ctenthex][\Cscr]{}%
\protect\end{array}\protect\right.\protect\label{eq:Fourier-Prod%
uctConvolutionInverse}\protect\end{eqnarray}\setcounter{CEht}{10%
}where convolution in the ``original'' space takes the exact sam%
e form as (\protect\ref{eq:Fourier-ConvolutionFourier}): 
\setcounter{Ceqindent}{0}\protect\begin{eqnarray}\protect\left.%
\protect\begin{array}{rcl}\protect\displaystyle\hspace{-1.3ex}&%
\protect\displaystyle(f\!\ast\!g)_n\equiv\mbox{$\protect
\displaystyle\protect\frac{1}{\sqrt N}$}\sum_{n^{\prime\!}=1}^Nf%
_{n'}g_{n-n^{\prime\!}}.\setlength{\Cscr}{\value{CEht}\Ctenthex}%
\addtolength{\Cscr}{-1.0ex}\protect\raisebox{0ex}[\value{CEht}%
\Ctenthex][\Cscr]{}\protect\end{array}\protect\right.\protect
\label{eq:Fourier-ConvolutionOriginal}\protect\end{eqnarray}%
\setcounter{CEht}{10}\par Now, the reader with a careful eye may 
have noticed that there is a slight problem with the definitions 
(\protect\ref{eq:Fourier-ConvolutionFourier}) and (\protect\ref{%
eq:Fourier-ConvolutionOriginal}) of the convolution operation. W%
e have specified, for definiteness, that the indices $n$ and $b$ 
will each take on values from $1$ to $N$, for the purposes of th%
is section. The indices in (\protect\ref{eq:Fourier-ConvolutionF%
ourier}) and (\protect\ref{eq:Fourier-ConvolutionOriginal}), how%
ever, take on values that clearly lie outside this range. For ex%
ample, consider the computation of the \raisebox{0ex}[1ex][0ex]{%
$\protect\displaystyle b=2$} value in the convolution (\protect
\ref{eq:Fourier-ConvolutionFourier}): \setcounter{Ceqindent}{0}%
\protect\begin{eqnarray}(F^{\:\!\!}\!\ast\!G)_2\hspace{-1.3ex}&%
\displaystyle=&\hspace{-1.3ex}\mbox{$\protect\displaystyle
\protect\frac{1}{\sqrt N}$}\sum_{b^{\prime\!}=1}^NF_{b'}G_{2-b^{%
\prime\!}}\protect\nonumber\setlength{\Cscr}{\value{CEht}%
\Ctenthex}\addtolength{\Cscr}{-1.0ex}\protect\raisebox{0ex}[%
\value{CEht}\Ctenthex][\Cscr]{}\\*[0ex]\protect\displaystyle
\hspace{-1.3ex}&\displaystyle=&\hspace{-1.3ex}\mbox{$\protect
\displaystyle\protect\frac{1}{\sqrt N}$}\setcounter{Cbscurr}{20}%
\setlength{\Cscr}{\value{Cbscurr}\Ctenthex}\addtolength{\Cscr}{-%
1.0ex}\protect\raisebox{0ex}[\value{Cbscurr}\Ctenthex][\Cscr]{}%
\hspace{-0ex}{\protect\left\{\setlength{\Cscr}{\value{Cbscurr}%
\Ctenthex}\addtolength{\Cscr}{-1.0ex}\protect\raisebox{0ex}[%
\value{Cbscurr}\Ctenthex][\Cscr]{}\protect\right.}\hspace{-0.25e%
x}\setlength{\Cscr}{\value{Cbscurr}\Ctenthex}\addtolength{\Cscr}%
{-1.0ex}\protect\raisebox{0ex}[\value{Cbscurr}\Ctenthex][\Cscr]{%
}\setcounter{CbsD}{\value{CbsC}}\setcounter{CbsC}{\value{CbsB}}%
\setcounter{CbsB}{\value{CbsA}}\setcounter{CbsA}{\value{Cbscurr}%
}F_1G_1+F_2G_0+F_3G_{\!-\:\!\!1}+\ldots+F_{\!N\:\!\!-1}G_{\!-\:%
\!\!N\:\!\!+3}+F_{\!N}G_{\!-\:\!\!N\:\!\!+2}\setlength{\Cscr}{%
\value{CbsA}\Ctenthex}\addtolength{\Cscr}{-1.0ex}\protect
\raisebox{0ex}[\value{CbsA}\Ctenthex][\Cscr]{}\hspace{-0.25ex}{%
\protect\left.\setlength{\Cscr}{\value{CbsA}\Ctenthex}%
\addtolength{\Cscr}{-1.0ex}\protect\raisebox{0ex}[\value{CbsA}%
\Ctenthex][\Cscr]{}\protect\right\}}\hspace{-0ex}\setlength{\Cscr
}{\value{CbsA}\Ctenthex}\addtolength{\Cscr}{-1.0ex}\protect
\raisebox{0ex}[\value{CbsA}\Ctenthex][\Cscr]{}\setcounter{CbsA}{%
\value{CbsB}}\setcounter{CbsB}{\value{CbsC}}\setcounter{CbsC}{%
\value{CbsD}}\setcounter{CbsD}{1}.\protect\nonumber\setlength{%
\Cscr}{\value{CEht}\Ctenthex}\addtolength{\Cscr}{-1.0ex}\protect
\raisebox{0ex}[\value{CEht}\Ctenthex][\Cscr]{}\protect\end{eqnar%
ray}\setcounter{CEht}{10}The index on $G$ should run from $1$ to 
$N$, like that on $F$; but here it is running from \raisebox{0ex%
}[1ex][0ex]{$\protect\displaystyle-N\:\!\!+2$} to $1$. So what h%
appened?\par The answer is that we made a rather uncritical use 
of the result (\protect\ref{eq:Fourier-KroneckerIdentity}) in de%
riving the convolution results (\protect\ref{eq:Fourier-ProductC%
onvolution}) and (\protect\ref{eq:Fourier-ProductConvolutionInve%
rse}). The correct generalisation of (\protect\ref{eq:Fourier-Kr%
oneckerIdentity}) is actually the following: \addtocounter{equat%
ion}{-1}\renewcommand\theequation{\protect\ref{eq:Fourier-Kronec%
kerIdentity}$'$}\setcounter{Ceqindent}{0}\protect\begin{eqnarray%
}\protect\left.\protect\begin{array}{rcl}\protect\displaystyle
\hspace{-1.3ex}&\protect\displaystyle\sum_{b=1}^Ne^{2\pi ibc/N\!%
}=\left\{\begin{array}{ll}N&\mbox{if \raisebox{0ex}[1ex][0ex]{$%
\protect\displaystyle c\equiv0$} (mod $N$),}\\\,0&\mbox{if 
\raisebox{0ex}[1ex][0ex]{$\protect\displaystyle c\not\equiv0$} (%
mod $N$),}\\\end{array}\right.\setlength{\Cscr}{\value{CEht}%
\Ctenthex}\addtolength{\Cscr}{-1.0ex}\protect\raisebox{0ex}[%
\value{CEht}\Ctenthex][\Cscr]{}\protect\end{array}\protect\right
.\protect\end{eqnarray}\renewcommand\theequation{\arabic{equatio%
n}}\setcounter{CEht}{10}where $c$ is an integer. In the case of 
the identity (\protect\ref{eq:Fourier-KroneckerIdentity}), we ha%
d \raisebox{0ex}[1ex][0ex]{$\protect\displaystyle c=n-n^{\prime
\!}$}, so that the condition that selects the nonzero term is 
\raisebox{0ex}[1ex][0ex]{$\protect\displaystyle n-n^{\prime\!}%
\equiv0\mbox{ (mod $N$)}$}, namely, \raisebox{0ex}[1ex][0ex]{$%
\protect\displaystyle n\equiv n^{\prime\!}\mbox{ (mod $N$)}$}. B%
ut since both $n$ and \raisebox{0ex}[1ex][0ex]{$\protect
\displaystyle n^{\prime\!}$} took on values between $1$ and $N$, 
the only way that this congruence could hold was for the case of 
\raisebox{0ex}[1ex][0ex]{$\protect\displaystyle n=n'$}; hence th%
e result (\protect\ref{eq:Fourier-KroneckerIdentity}) was, in fa%
ct, correct. For the final sum in (\protect\ref{eq:Fourier-Final%
LineConvolution}), on the other hand, the correct application of 
(\protect\ref{eq:Fourier-KroneckerIdentity}$'$) yields 
\setcounter{Ceqindent}{0}\protect\begin{eqnarray}\hspace{-1.3ex}%
&\displaystyle\sum_{n=1}^Ne^{2\pi i(b^{\prime\!}+b^{\prime\:\!\!%
\prime\!}-b)n/N\!}=\left\{\begin{array}{ll}N&\mbox{if \raisebox{%
0ex}[1ex][0ex]{$\protect\displaystyle b^{\prime^{\:\!\!}\prime\!%
}\equiv b-b^{\prime\!}$} (mod $N$),}\\\,0&\mbox{if \raisebox{0ex%
}[1ex][0ex]{$\protect\displaystyle b^{\prime^{\:\!\!}\prime\!}%
\not\equiv b-b^{\prime\!}$} (mod $N$).}\\\end{array}\right.%
\protect\nonumber\setlength{\Cscr}{\value{CEht}\Ctenthex}%
\addtolength{\Cscr}{-1.0ex}\protect\raisebox{0ex}[\value{CEht}%
\Ctenthex][\Cscr]{}\protect\end{eqnarray}\setcounter{CEht}{10}In 
other words, we should not have selected out the term for which 
\raisebox{0ex}[1ex][0ex]{$\protect\displaystyle b^{\prime^{\:\!%
\!}\prime\!}=b-b^{\prime\!}$} (which may, in general, fall outsi%
de the range of values taken on by indices $b$), but rather we s%
hould have looked for that (permissible) value of \raisebox{0ex}%
[1ex][0ex]{$\protect\displaystyle b^{\prime^{\:\!\!}\prime\!}$} 
that was congruent to \raisebox{0ex}[1ex][0ex]{$\protect
\displaystyle b-b^{\prime\!}$} modulo~$N$. For some values of $b%
$ and \raisebox{0ex}[1ex][0ex]{$\protect\displaystyle b^{\prime
\!}$}, the permissible value of \raisebox{0ex}[1ex][0ex]{$%
\protect\displaystyle b^{\prime\:\!\!\prime\!}$} will simply be 
\raisebox{0ex}[1ex][0ex]{$\protect\displaystyle b-b'$}; for othe%
rs, it will be \raisebox{0ex}[1ex][0ex]{$\protect\displaystyle N%
\:\!\!+b-b'$}. If we had chosen a different range of values for 
$b$, then other multiples of $N$ would have to be added in order 
to obtain values of \raisebox{0ex}[1ex][0ex]{$\protect
\displaystyle b^{\prime\:\!\!\prime\!}$} within its allowed rang%
e.\par Catering for these various cases in the mathematical defi%
nition of the convolution is a severe inconvenience, and spoils 
the elegance of its expression. The usual alternative is to deem 
that all indices $n$ and $b$ are, in fact, permitted to take on 
\mbox{}\protect\/{\protect\em any\protect\/} integral values, bu%
t that it is implicitly assumed that any index arithmetic is ult%
imately carried out modulo $N$, and then shifted into the chosen 
range of values for indices of that type. This effectively stipu%
lates that the lattices in $n$-space and $b$-space are to be und%
erstood to be formed into rings of $N$ sites each.\par Whilst th%
is prescription provides a mathematical solution, its physical r%
amifications are both subtle and potentially devastating. Indeed%
, it is this ``trick'' that is fundamentally responsible for lar%
gely hiding the aliasing problem that has affected the fidelity 
of lattice calculations in the past.\par\refstepcounter{section}%
\vspace{1.5\baselineskip}\par{\centering\bf\thesection. Employin%
g lattices for physical applications\\*[0.5\baselineskip]}%
\protect\indent\label{sect:Physical}The previous section has rev%
iewed the fundamental elements of Fourier theory in purely mathe%
matical terms, without reference to any physical application or 
interpretation of the results. We shall now review how correspon%
dences between these mathematical constructs and physical realit%
y are generally constructed.\par The lattice of sites in the ``o%
riginal'' $n$-space is usually taken to correspond to equispaced 
positions in position space; namely, we deem that the position $%
x_n$ corresponding to the lattice site $n$ is just \setcounter{C%
eqindent}{0}\protect\begin{eqnarray}\protect\left.\protect\begin
{array}{rcl}\protect\displaystyle\hspace{-1.3ex}&\protect
\displaystyle x_{n\!}\equiv na,\setlength{\Cscr}{\value{CEht}%
\Ctenthex}\addtolength{\Cscr}{-1.0ex}\protect\raisebox{0ex}[%
\value{CEht}\Ctenthex][\Cscr]{}\protect\end{array}\protect\right
.\protect\label{eq:Physical-DefineXn}\protect\end{eqnarray}%
\setcounter{CEht}{10}where $a$ is the ``lattice spacing''\mbox{$%
\!$}, namely, the real (dimensionful) distance between adjacent 
lattice sites. (Recall that we are restricting our attention her%
e to one-dimensional lattices; the generalisation to an arbitrar%
y number of dimensions is elementary.) The lattice has a total (%
real) length of \setcounter{Ceqindent}{0}\protect\begin{eqnarray%
}\protect\left.\protect\begin{array}{rcl}\protect\displaystyle
\hspace{-1.3ex}&\protect\displaystyle L=N^{\!}a.\setlength{\Cscr
}{\value{CEht}\Ctenthex}\addtolength{\Cscr}{-1.0ex}\protect
\raisebox{0ex}[\value{CEht}\Ctenthex][\Cscr]{}\protect\end{array%
}\protect\right.\protect\label{eq:Physical-Length}\protect\end{e%
qnarray}\setcounter{CEht}{10}\par The relation (\protect\ref{eq:%
 Fourier-InverseFourier}) now tells us that \mbox{}\protect\/{%
\protect\em any\protect\/} arbitrary function \raisebox{0ex}[1ex%
][0ex]{$\protect\displaystyle f_{\:\!\!n}$} defined on the $n$ s%
ites $x_n$ can be written as a suitable linear superposition of 
$N$ complex exponential ``waves''\mbox{$\!$}. Based on our exper%
ience with quantum mechanics, we interpret each such wave as bei%
ng an ``eigenstate of momentum''\mbox{$\!$}. Now, inverting (%
\protect\ref{eq:Physical-DefineXn}) for $n$, we have \setcounter
{Ceqindent}{0}\protect\begin{eqnarray}\hspace{-1.3ex}&%
\displaystyle e^{2\pi ibn/N\!}=e^{2\pi ibx_{\:\!\!n\:\!\!}/N\:\!%
\!a\!}.\protect\nonumber\setlength{\Cscr}{\value{CEht}\Ctenthex}%
\addtolength{\Cscr}{-1.0ex}\protect\raisebox{0ex}[\value{CEht}%
\Ctenthex][\Cscr]{}\protect\end{eqnarray}\setcounter{CEht}{10}Fo%
r this to be equivalent to the oscillating part of our usual def%
inition of a plane wave eigenstate of momentum $p$, namely, 
\raisebox{0ex}[1ex][0ex]{$\protect\displaystyle e^{ipx\!}$} (whe%
re I shall always use units in which \raisebox{0ex}[1ex][0ex]{$%
\protect\displaystyle\hbar=1$}), we therefore require \setcounter
{Ceqindent}{0}\protect\begin{eqnarray}\protect\left.\protect
\begin{array}{rcl}\protect\displaystyle\hspace{-1.3ex}&\protect
\displaystyle p_b\equiv\mbox{$\protect\displaystyle\protect\frac
{2\pi}{N\:\!\!a}$}^{\:\!}b.\setlength{\Cscr}{\value{CEht}%
\Ctenthex}\addtolength{\Cscr}{-1.0ex}\protect\raisebox{0ex}[%
\value{CEht}\Ctenthex][\Cscr]{}\protect\end{array}\protect\right
.\protect\label{eq:Physical-DefineP}\protect\end{eqnarray}%
\setcounter{CEht}{10}In other words, the momentum value $p$ is s%
imply proportional to the index $b$. Now, since the ``length'' o%
f the lattice in $b$-space is also $N$ sites, the relationship (%
\protect\ref{eq:Physical-DefineP}) tells us that the ``length'' 
of the lattice in momentum space is just \setcounter{Ceqindent}{%
0}\protect\begin{eqnarray}\protect\left.\protect\begin{array}{rc%
l}\protect\displaystyle\hspace{-1.3ex}&\protect\displaystyle{%
\protect\it\Lambda\!\:}\equiv\mbox{$\protect\displaystyle\protect
\frac{2\pi}{a}$}.\setlength{\Cscr}{\value{CEht}\Ctenthex}%
\addtolength{\Cscr}{-1.0ex}\protect\raisebox{0ex}[\value{CEht}%
\Ctenthex][\Cscr]{}\protect\end{array}\protect\right.\protect
\label{eq:Physical-DefineLambda}\protect\end{eqnarray}\setcounter
{CEht}{10}(The somewhat annoying factors of $2\pi$ in (\protect
\ref{eq:Physical-DefineP}) and (\protect\ref{eq:Physical-DefineL%
ambda}), that float around all Fourier expressions involving $p$%
, fundamentally arise because we have chosen our units such that 
\raisebox{0ex}[1ex][0ex]{$\protect\displaystyle\hbar=1$} rather 
than \raisebox{0ex}[1ex][0ex]{$\protect\displaystyle h=1$}. Engi%
neers essentially make the alternative choice, so that their ``m%
omentum space'' variable is not complicated in this way. However%
, this throws an extra factor of $2\pi$ into the Fourier transfo%
rm of the derivative operator, and so is inconvenient when we wa%
nt to deal extensively with the formulation of differential equa%
tions, as we do in quantum field theory.)\par So far, our connec%
tion of $n$-space to position space, and $b$-space to momentum s%
pace, has retained the fundamental mathematical symmetry between 
the two spaces that was evident in \mbox{Sec.~$\:\!\!$}\protect
\ref{sect:Fourier}. However, once we start thinking about these 
spaces with a physical interpretation in mind, this mathematical 
symmetry begins to dissolve, because position space and momentum 
space do not enter into the laws of physics in identical ways---%
far from it, in fact.\par Consider the situation in position spa%
ce. Our fundamental laws of physics are translationally invarian%
t, which means that no particular position $x$ is fundamentally 
different than any other particular position \raisebox{0ex}[1ex]%
[0ex]{$\protect\displaystyle{x}{}^{\raisebox{-0.25ex}{$%
\scriptstyle{\prime\!}$}}$}. This translational invariance is no%
t obeyed by the position-space lattice, as we have defined it so 
far: it has a finite length $L$, and hence has two ``ends'' whic%
h have only one neighbouring lattice site each. In this context, 
we willingly embrace the ``ring'' interpretation of the index $n%
$ described in the previous section, so that position space is e%
ffectively wrapped into a ring of $N$ lattice sites. This ensure%
s that there are no ``ends''\mbox{$\!$}, albeit at the expense o%
f imposing a fundamental periodicity on position space---namely, 
the variable $x$ can be defined to run from \raisebox{0ex}[1ex][%
0ex]{$\protect\displaystyle x=-\infty$} to \raisebox{0ex}[1ex][0%
ex]{$\protect\displaystyle x=+\infty$}, but every function of $x%
$ is forced to repeat identically with a periodicity of $L$. Our 
gut physical instinct tells us that, if we can make $L$ sufficie%
ntly large for the physical system we are wishing to model, the 
``finite volume'' artefacts of this periodicity (\mbox{}\protect
\/{\protect\em i.e.\protect\/}, the interactions between one ``c%
opy'' of the physical system in position space and the adjacent 
``copies'' of itself) can be made sufficiently small that they d%
o not detract substantially from the fidelity of the mathematica%
l model we are employing.\par The situation in momentum space is 
completely different. Here we again have a lattice of finite ``l%
ength''\mbox{$\!$}\ ${\protect\it\Lambda\!\:}$. However, we do n%
ot have any fundamental property of ``momentum translation invar%
iance'' that we wish to maintain. (The relativity of motion of G%
alilean or Lorentzian kinematics is somewhat more subtle than a 
simple translation in momentum space, and we can generally perfo%
rm a transformation to the ``centre\ of momentum'' frame of refe%
rence, so that such invariance is of no practical concern anyway%
.) Rather, we consider the momentum ``length''\mbox{$\!$}\ ${%
\protect\it\Lambda\!\:}$ to provide an upper limit to the moment%
um states that can contribute to our calculations. Indeed, since 
we have (in any sensible calculation) made the abovementioned tr%
ansformation to a suitable ``centre\ of momentum'' frame of refe%
rence, we make use of our freedom to choose the range of $b$ (an%
d hence $p$) to ensure that momentum space is ``centred'' on 
\raisebox{0ex}[1ex][0ex]{$\protect\displaystyle p=0$}; in other 
words, we choose to have $p$ falling in the range \raisebox{0ex}%
[1ex][0ex]{$\protect\displaystyle-\:\!\!{\protect\it\Lambda\!\:}%
<p\leq+\:\!\!{\protect\it\Lambda\!\:}$}. In the context of quant%
um field theory, this ``cutoff'' in momentum space is actually q%
uite convenient, regulating the formalism automatically by remov%
ing all of the higher-momentum states from our calculations.\par
Again, our very language here demonstrates the fundamentally dif%
ferent way that we view momentum space compared to position spac%
e. We accept implicitly the discussion of a ``momentum state''%
\mbox{$\!$}, because it is something that we are used to conside%
ring, even in undergraduate physics. Mathematically, however, a 
``momentum state'' is actually quite a singular object. It corre%
sponds to a function that has but a single component in momentum 
space. For example, if we are considering the momentum state cor%
responding to the momentum-space index \raisebox{0ex}[1ex][0ex]{%
$\protect\displaystyle b=3$}, then the function in momentum spac%
e takes the form \setcounter{Ceqindent}{0}\protect\begin{eqnarra%
y}\protect\left.\protect\begin{array}{rcl}\protect\displaystyle
\hspace{-1.3ex}&\protect\displaystyle F_b=A^{\,}\delta_{b3},%
\setlength{\Cscr}{\value{CEht}\Ctenthex}\addtolength{\Cscr}{-1.0%
ex}\protect\raisebox{0ex}[\value{CEht}\Ctenthex][\Cscr]{}\protect
\end{array}\protect\right.\protect\label{eq:Physical-MomentumSta%
teExample}\protect\end{eqnarray}\setcounter{CEht}{10}where $A$ i%
s some arbitrary amplitude. The reason that we accept such a fun%
ction, without need for explicit comment, is that we have been l%
ong accustomed to the fact that any translationally-invariant wa%
ve equation will have eigenstates that are of precisely this for%
m. Indeed, an engineer would equally accept a plane wave functio%
n without argument---perhaps, at worst, insisting on a real wave%
, such as \setcounter{Ceqindent}{0}\protect\begin{eqnarray}%
\protect\left.\protect\begin{array}{rcl}\protect\displaystyle
\hspace{-1.3ex}&\protect\displaystyle F_b=\mbox{$\protect
\displaystyle\protect\frac{A}{2}$}\setcounter{Cbscurr}{20}%
\setlength{\Cscr}{\value{Cbscurr}\Ctenthex}\addtolength{\Cscr}{-%
1.0ex}\protect\raisebox{0ex}[\value{Cbscurr}\Ctenthex][\Cscr]{}%
\hspace{-0ex}{\protect\left\{\setlength{\Cscr}{\value{Cbscurr}%
\Ctenthex}\addtolength{\Cscr}{-1.0ex}\protect\raisebox{0ex}[%
\value{Cbscurr}\Ctenthex][\Cscr]{}\protect\right.}\hspace{-0.25e%
x}\setlength{\Cscr}{\value{Cbscurr}\Ctenthex}\addtolength{\Cscr}%
{-1.0ex}\protect\raisebox{0ex}[\value{Cbscurr}\Ctenthex][\Cscr]{%
}\setcounter{CbsD}{\value{CbsC}}\setcounter{CbsC}{\value{CbsB}}%
\setcounter{CbsB}{\value{CbsA}}\setcounter{CbsA}{\value{Cbscurr}%
}\delta_{b3}+\delta_{b,-3}\setlength{\Cscr}{\value{CbsA}\Ctenthex
}\addtolength{\Cscr}{-1.0ex}\protect\raisebox{0ex}[\value{CbsA}%
\Ctenthex][\Cscr]{}\hspace{-0.25ex}{\protect\left.\setlength{%
\Cscr}{\value{CbsA}\Ctenthex}\addtolength{\Cscr}{-1.0ex}\protect
\raisebox{0ex}[\value{CbsA}\Ctenthex][\Cscr]{}\protect\right\}}%
\hspace{-0ex}\setlength{\Cscr}{\value{CbsA}\Ctenthex}\addtolength
{\Cscr}{-1.0ex}\protect\raisebox{0ex}[\value{CbsA}\Ctenthex][%
\Cscr]{}\setcounter{CbsA}{\value{CbsB}}\setcounter{CbsB}{\value{%
CbsC}}\setcounter{CbsC}{\value{CbsD}}\setcounter{CbsD}{1}.%
\setlength{\Cscr}{\value{CEht}\Ctenthex}\addtolength{\Cscr}{-1.0%
ex}\protect\raisebox{0ex}[\value{CEht}\Ctenthex][\Cscr]{}\protect
\end{array}\protect\right.\protect\label{eq:Physical-MomentumSta%
teReal}\protect\end{eqnarray}\setcounter{CEht}{10}In contrast, w%
e wouldn't generally talk about a ``position state''; such a phr%
ase would need to be explained in order to be intelligible. Howe%
ver, if we were to write down the position-space analogue\ of (%
\protect\ref{eq:Physical-MomentumStateExample}), namely, 
\setcounter{Ceqindent}{0}\protect\begin{eqnarray}\hspace{-1.3ex}%
&\displaystyle F_n=A^{\,}\delta_{n3},\protect\nonumber\setlength
{\Cscr}{\value{CEht}\Ctenthex}\addtolength{\Cscr}{-1.0ex}\protect
\raisebox{0ex}[\value{CEht}\Ctenthex][\Cscr]{}\protect\end{eqnar%
ray}\setcounter{CEht}{10}then we would immediately recognise wha%
t is being considered: a delta-function in position space. Why t%
he different description? Simply because such a function is \mbox
{}\protect\/{\protect\em not\protect\/} an eigenstate of any nat%
urally-occurring equation of physics that we are used to conside%
ring. Rather, it is a fairly advanced mathematical construct tha%
t is used in more abstract analyses of physical systems. The eng%
ineer likewise accepts it as such; but in this case there is no 
objection to its form---no insistence that it be ``symmetrised'' 
into a form similar to (\protect\ref{eq:Physical-MomentumStateRe%
al}). Why not? Simply because it is in position space---``real'' 
space---that the engineer insists on quantities being real (and 
indeed the physicist concurs, with the refinement that only \mbox
{}\protect\/{\protect\em physically observable\protect\/} quanti%
ties need actually be real in position space). Momentum space, i%
n contrast, is an ``artificial'' or ``mathematical'' space, with 
which we learn to make an intellectual connection, but which is 
not the ``real'' world of our physical senses. Indeed, for a fun%
ction to be real in position space, its Fourier transform needs 
to be \mbox{}\protect\/{\protect\em Hermitian\protect\/} (not re%
al) in momentum space, namely, \setcounter{Ceqindent}{0}\protect
\begin{eqnarray}\protect\left.\protect\begin{array}{rcl}\protect
\displaystyle\hspace{-1.3ex}&\protect\displaystyle F_{\:\!\!\!-b%
}^{}=F^{\ast\!}_b,\setlength{\Cscr}{\value{CEht}\Ctenthex}%
\addtolength{\Cscr}{-1.0ex}\protect\raisebox{0ex}[\value{CEht}%
\Ctenthex][\Cscr]{}\protect\end{array}\protect\right.\protect
\label{eq:Physical-HermitianMomentumSpace}\protect\end{eqnarray}%
\setcounter{CEht}{10}where the index $-b$ is understood to be ta%
ken modulo \raisebox{0ex}[1ex][0ex]{$\protect\displaystyle N$} a%
nd shifted into the correct range for momentum indices (although 
if we \mbox{}\protect\/{\protect\em have\protect\/} centred\ the 
momentum index range on \raisebox{0ex}[1ex][0ex]{$\protect
\displaystyle p=0$}, then this caveat is not required). If we we%
re to furthermore insist that the function in momentum space be 
\mbox{}\protect\/{\protect\em real\protect\/}, then we would be 
restricting the function in position space to be an even functio%
n of $x$, which is not appropriate or applicable in general. Thu%
s, the physicist and the engineer both accept, without question, 
that functions in momentum space will, in general, be complex-va%
lued, even when they are describing real functions in position s%
pace.\par Now, let us leave the divergence between our physical 
interpretation of position space and momentum space to one side, 
for the moment. Instead, let us focus on a more fundamental ques%
tion: how do we connect the continuum spaces of the real world w%
ith the discrete spaces of the lattice world?\par The simple ans%
wer is that we take the number of lattice sites, $N$, to infinit%
y. However, by itself, simply taking \raisebox{0ex}[1ex][0ex]{$%
\protect\displaystyle N\rightarrow\infty$} does not provide a de%
finite connection with the real world. Rather, we must also spec%
ify how the lattice spacing, $a$, is to behave as \raisebox{0ex}%
[1ex][0ex]{$\protect\displaystyle N\rightarrow\infty$}. Now, as 
a general mathematical proposition, there is of course an infini%
te number of ways in which we can make $a$ depend functionally o%
n $N$. However, there are three particularly simple choices of $%
a(N)$ that allow a direct physical interpretation, and are of ge%
neral importance in all practical applications of Fourier theory%
. I will now discuss each of these in turn, and review explicitl%
y how the finite-lattice results of \mbox{Sec.~$\:\!\!$}\protect
\ref{sect:Fourier} are transformed into their infinite-lattice c%
ounterparts.\par The simplest choice of $a(N)$ is just 
\setcounter{Ceqindent}{0}\protect\begin{eqnarray}\protect\left.%
\protect\begin{array}{rcl}\protect\displaystyle\hspace{-1.3ex}&%
\protect\displaystyle a(N)=a=\mbox{constant}.\setlength{\Cscr}{%
\value{CEht}\Ctenthex}\addtolength{\Cscr}{-1.0ex}\protect
\raisebox{0ex}[\value{CEht}\Ctenthex][\Cscr]{}\protect\end{array%
}\protect\right.\protect\label{eq:Physical-ConstantA}\protect\end
{eqnarray}\setcounter{CEht}{10}In other words, the real distance 
between lattice sites is kept constant; as we increase $N$, the 
extra sites simply increase the length of the lattice linearly, 
as shown by (\protect\ref{eq:Physical-Length}): \setcounter{Ceqi%
ndent}{0}\protect\begin{eqnarray}\protect\left.\protect\begin{ar%
ray}{rcl}\protect\displaystyle\hspace{-1.3ex}&\protect
\displaystyle L(N)=N^{\!}a,\hspace{6ex}\mbox{\mbox{}\protect\/{%
\protect\em i.e.\protect\/}, }\hspace{2ex}L(N)\propto N.%
\setlength{\Cscr}{\value{CEht}\Ctenthex}\addtolength{\Cscr}{-1.0%
ex}\protect\raisebox{0ex}[\value{CEht}\Ctenthex][\Cscr]{}\protect
\end{array}\protect\right.\protect\label{eq:Physical-LengthConst%
antA}\protect\end{eqnarray}\setcounter{CEht}{10}The constant spa%
cing $a$ remains a part of the mathematical description in the c%
ontinuum limit \raisebox{0ex}[1ex][0ex]{$\protect\displaystyle N%
\rightarrow\infty$}, and hence must correspond to some physical 
property of the system being modelled; in other words, $a$ is a 
\mbox{}\protect\/{\protect\em physically real\protect\/} quantit%
y, rather than just a mathematical symbol of our intermediate ca%
lculations. In practice, the prescription (\protect\ref{eq:Physi%
cal-ConstantA}) would only be of use for systems that do, physic%
ally, have a lattice structure of periodicity $a$, such as a cry%
stal.\par\mbox{Eq.~\raisebox{0ex}[1ex][0ex]{$\protect
\displaystyle\!$}}(\protect\ref{eq:Physical-DefineLambda}) tells 
us that the ``length'' of the momentum-space lattice will be a c%
onstant, regardless of the value of $N$: \setcounter{Ceqindent}{%
0}\protect\begin{eqnarray}\protect\left.\protect\begin{array}{rc%
l}\protect\displaystyle\hspace{-1.3ex}&\protect\displaystyle{%
\protect\it\Lambda\!\:}(N)=\mbox{$\protect\displaystyle\protect
\frac{2\pi}{a}$}=\mbox{constant}.\setlength{\Cscr}{\value{CEht}%
\Ctenthex}\addtolength{\Cscr}{-1.0ex}\protect\raisebox{0ex}[%
\value{CEht}\Ctenthex][\Cscr]{}\protect\end{array}\protect\right
.\protect\label{eq:Physical-LambdaConstantA}\protect\end{eqnarra%
y}\setcounter{CEht}{10}This implies that, as $N$ increases, more 
and more momentum-space lattice sites will be packed into this f%
ixed ``length'' of ${\protect\it\Lambda\!\:}$; \mbox{Eq.~%
\raisebox{0ex}[1ex][0ex]{$\protect\displaystyle\!$}}(\protect\ref
{eq:Physical-DefineP}) tells us that the ``spacing'' in momentum 
space between adjacent lattice sites is \setcounter{Ceqindent}{0%
}\protect\begin{eqnarray}\protect\left.\protect\begin{array}{rcl%
}\protect\displaystyle\hspace{-1.3ex}&\protect\displaystyle\delta
^{\:\!\!}p=\mbox{$\protect\displaystyle\protect\frac{2\pi}{N\:\!%
\!a}$}.\setlength{\Cscr}{\value{CEht}\Ctenthex}\addtolength{\Cscr
}{-1.0ex}\protect\raisebox{0ex}[\value{CEht}\Ctenthex][\Cscr]{}%
\protect\end{array}\protect\right.\protect\label{eq:Physical-Del%
taPConstantA}\protect\end{eqnarray}\setcounter{CEht}{10}Hence, i%
n the limit \raisebox{0ex}[1ex][0ex]{$\protect\displaystyle N%
\rightarrow\infty$}, momentum space will become a continuum, wit%
hin the ``length''\mbox{$\!$}\ \raisebox{0ex}[1ex][0ex]{$\protect
\displaystyle{\protect\it\Lambda\!\:}=2\pi/a$}. This is the ``fi%
rst Brillouin zone'' of the crystallographer.\par We must now de%
termine how the Fourier transform expressions of \mbox{Sec.~$\:%
\!\!$}\protect\ref{sect:Fourier} are to be transformed in the li%
mit \raisebox{0ex}[1ex][0ex]{$\protect\displaystyle N\rightarrow
\infty$}. Clearly, sums in position space will continue to be su%
ms, but now over the infinite number of lattice sites. It is usu%
al to first change the range of $n$ from the $1$ to $N$ of \mbox
{Sec.~$\:\!\!$}\protect\ref{sect:Fourier} to a more ``symmetrica%
l'' definition, such as from the least integer $>-N/2$ to the gr%
eatest integer $\leq+N/2$, so that in the limit \raisebox{0ex}[1%
ex][0ex]{$\protect\displaystyle N\rightarrow\infty$} the sum ove%
r $n$ will be from $-\infty$ to $+\infty$. (The alternative woul%
d be to have an infinite lattice that possesses a single end, li%
ke a ray in geometry; this somewhat complicates the mathematical 
description, because instead of the continuous Fourier transform 
we would end up with the Laplace transform, which is not quite a%
s convenient for our purposes.)\par On the other hand, sums in m%
omentum space will clearly be converted into integrals, over the 
continuum of the first Brillouin zone. We have to be careful, ho%
wever, in establishing this correspondence. To illustrate why th%
is requires extreme care, it is useful to first examine the defi%
nition of the convolution in $b$-space, namely, \mbox{Eq.~%
\raisebox{0ex}[1ex][0ex]{$\protect\displaystyle\!$}}(\protect\ref
{eq:Fourier-ConvolutionFourier}). The usual transition to the co%
ntinuum limit entails the insertion of a factor of \raisebox{0ex%
}[1ex][0ex]{$\protect\displaystyle\delta^{\:\!\!}p$} inside the 
sum, balanced by a compensating factor of \raisebox{0ex}[1ex][0e%
x]{$\protect\displaystyle\delta^{\:\!\!}p$} outside it: 
\setcounter{Ceqindent}{0}\protect\begin{eqnarray}\protect\left.%
\protect\begin{array}{rcl}\protect\displaystyle\hspace{-1.3ex}&%
\protect\displaystyle(F^{\:\!\!}\!\ast\!G)_b=\mbox{$\protect
\displaystyle\protect\frac{1}{\delta^{\:\!\!}p\:\!\sqrt N}$}\sum
_{b^{\prime\!}}\:\!\!\delta^{\:\!\!}p\,F_{b'}G_{b-b^{\prime\!}}=%
\mbox{$\protect\displaystyle\protect\frac{a\sqrt N}{2\pi}$}^{\!}%
\sum_{b^{\prime\!}}\:\!\!\delta^{\:\!\!}p\,F_{b'}G_{b-b^{\prime
\!}},\setlength{\Cscr}{\value{CEht}\Ctenthex}\addtolength{\Cscr}%
{-1.0ex}\protect\raisebox{0ex}[\value{CEht}\Ctenthex][\Cscr]{}%
\protect\end{array}\protect\right.\protect\label{eq:Physical-Con%
vFixedAFirst}\protect\end{eqnarray}\setcounter{CEht}{10}where it 
is understood that the sum over \raisebox{0ex}[1ex][0ex]{$%
\protect\displaystyle b^{\prime\!}$} is over the range of intege%
rs of the first Brillouin zone. However, if we now take the limi%
t \raisebox{0ex}[1ex][0ex]{$\protect\displaystyle N\rightarrow
\infty$}, we do not obtain the result we might expect: 
\setcounter{Ceqindent}{0}\protect\begin{eqnarray}\protect\left.%
\protect\begin{array}{rcl}\protect\displaystyle\hspace{-1.3ex}&%
\protect\displaystyle\lim_{N\rightarrow\infty}\!\mbox{$\protect
\displaystyle\protect\frac{a\sqrt N}{2\pi}$}^{\!}\sum_{b^{\prime
\!}}\:\!\!\delta^{\:\!\!}p\,F_{b'}G_{b-b^{\prime\!}}=\lim_{N%
\rightarrow\infty}\!\mbox{$\protect\displaystyle\protect\frac{a%
\sqrt N}{2\pi}$}^{\!}\hspace{-0.5mm}{\protect\mbox{}}\hspace{-0.%
1mm}\protect\int{\protect\mbox{}}\hspace{-0.5mm}\hspace{-0.6mm}d%
p'{}^{\:\!}F(p')^{\:\!}G(p\!-\!p')=\infty,\setlength{\Cscr}{%
\value{CEht}\Ctenthex}\addtolength{\Cscr}{-1.0ex}\protect
\raisebox{0ex}[\value{CEht}\Ctenthex][\Cscr]{}\protect\end{array%
}\protect\right.\protect\label{eq:Physical-InfiniteFixedA}%
\protect\end{eqnarray}\setcounter{CEht}{10}where we are using (%
\protect\ref{eq:Physical-DefineP}) to relate $b$ to $p$ (and 
\raisebox{0ex}[1ex][0ex]{$\protect\displaystyle b^{\prime\!}$} t%
o \raisebox{0ex}[1ex][0ex]{$\protect\displaystyle p'$}), and we 
are assuming that the integral of \raisebox{0ex}[1ex][0ex]{$%
\protect\displaystyle F(p')^{\:\!}G(p\!-\!p')$} is finite. The p%
roblem is the factor of $\sqrt N$ outside the integral, which di%
verges as \raisebox{0ex}[1ex][0ex]{$\protect\displaystyle N%
\rightarrow\infty$}.\par To avoid this fate, we need to redefine 
all functions in momentum space in the limit \raisebox{0ex}[1ex]%
[0ex]{$\protect\displaystyle N\rightarrow\infty$}, by multiplyin%
g them by a multiplicative factor that is a function of $N$. Thi%
s is essentially an elementary application of the idea of renorm%
alisation. Now, since $F(p)$ and $G(p)$ appear in the convolutio%
n operation symmetrically, we must apply the same renormalisatio%
n factor to each of them. However, the result of the convolution 
operation is itself a function in momentum space, and so should 
also be renormalised by the same factor. Taking these considerat%
ions into account, the divergence of (\protect\ref{eq:Physical-I%
nfiniteFixedA}) implies that, corresponding to any function $H(p%
)$ in momentum space, we should define a renormalised function 
\setcounter{Ceqindent}{0}\protect\begin{eqnarray}\protect\left.%
\protect\begin{array}{rcl}\protect\displaystyle\hspace{-1.3ex}&%
\protect\displaystyle\,\overline{\!H}(p)\equiv H(p)\sqrt N,%
\setlength{\Cscr}{\value{CEht}\Ctenthex}\addtolength{\Cscr}{-1.0%
ex}\protect\raisebox{0ex}[\value{CEht}\Ctenthex][\Cscr]{}\protect
\end{array}\protect\right.\protect\label{eq:Physical-Renormalise%
dConstantA}\protect\end{eqnarray}\setcounter{CEht}{10}so that 
\setcounter{Ceqindent}{0}\protect\begin{eqnarray}\protect\left.%
\protect\begin{array}{rcl}\protect\displaystyle\hspace{-1.3ex}&%
\protect\displaystyle H(p)=\mbox{$\protect\displaystyle\protect
\frac{\,\overline{\!H}(p)}{\sqrt N}$}.\setlength{\Cscr}{\value{C%
Eht}\Ctenthex}\addtolength{\Cscr}{-1.0ex}\protect\raisebox{0ex}[%
\value{CEht}\Ctenthex][\Cscr]{}\protect\end{array}\protect\right
.\protect\label{eq:Physical-InvertedRenormalisedConstantA}%
\protect\end{eqnarray}\setcounter{CEht}{10}If we insert these re%
normalised functions into (\protect\ref{eq:Physical-ConvFixedAFi%
rst}), we now obtain \setcounter{Ceqindent}{0}\protect\begin{eqn%
array}\hspace{-1.3ex}&\displaystyle\mbox{$\protect\displaystyle
\protect\frac{\overline{(F^{\:\!\!}\!\ast\!G)}_b}{\sqrt N}$}=%
\mbox{$\protect\displaystyle\protect\frac{1}{\delta^{\:\!\!}p\:%
\!\sqrt N}$}\sum_{b^{\prime\!}}\:\!\!\delta^{\:\!\!}p\,\mbox{$%
\protect\displaystyle\protect\frac{\,\overline{\!F}_{\!b'}}{\sqrt
N}$}\mbox{$\protect\displaystyle\protect\frac{\,\overline{\!G}_{%
\:\!\!b-b^{\prime\!}}}{\sqrt N}$},\protect\nonumber\setlength{%
\Cscr}{\value{CEht}\Ctenthex}\addtolength{\Cscr}{-1.0ex}\protect
\raisebox{0ex}[\value{CEht}\Ctenthex][\Cscr]{}\protect\end{eqnar%
ray}\setcounter{CEht}{10}so that \setcounter{Ceqindent}{0}%
\protect\begin{eqnarray}\hspace{-1.3ex}&\displaystyle\overline{(%
 F^{\:\!\!}\!\ast\!G)}_b=\mbox{$\protect\displaystyle\protect
\frac{1}{N^{\:\!}\delta^{\:\!\!}p}$}\sum_{b^{\prime\!}}\:\!\!%
\delta^{\:\!\!}p\,\,\overline{\!F}_{\!b'}\,\overline{\!G}_{\:\!%
\!b-b^{\prime\!}}=\mbox{$\protect\displaystyle\protect\frac{a}{2%
\pi}$}^{\!}\sum_{b^{\prime\!}}\:\!\!\delta p\,\,\overline{\!F}_{%
\!b'}\,\overline{\!G}_{\:\!\!b-b^{\prime\!}},\protect\nonumber
\setlength{\Cscr}{\value{CEht}\Ctenthex}\addtolength{\Cscr}{-1.0%
ex}\protect\raisebox{0ex}[\value{CEht}\Ctenthex][\Cscr]{}\protect
\end{eqnarray}\setcounter{CEht}{10}We can now take the \raisebox
{0ex}[1ex][0ex]{$\protect\displaystyle N\rightarrow\infty$} limi%
t without encountering any divergence problems: \setcounter{Ceqi%
ndent}{0}\protect\begin{eqnarray}\protect\left.\protect\begin{ar%
ray}{rcl}\protect\displaystyle\hspace{-1.3ex}&\protect
\displaystyle\overline{(F^{\:\!\!}\!\ast\!G)}(p)=\mbox{$\protect
\displaystyle\protect\frac{1}{{\protect\it\Lambda\!\:}^{\:\!}}$}%
^{\!}\hspace{-0.5mm}{\protect\mbox{}}\hspace{-0.1mm}\makebox[0ex%
]{\raisebox{-5.4mm}{\hspace{-0ex}\hspace{1mm}\makebox[0ex]{%
\scriptsize{\protect\mbox{}}${\protect\it\Lambda\!\:}$}}\hspace{%
0ex}\hspace{-1.5mm}}\protect\int{\protect\mbox{}}\hspace{-0.5mm}%
\hspace{-0.6mm}dp'{}^{\:\!}\,\overline{\!F}(p')^{\:\!}\,\overline
{\!G}(p\!-\!p'),\setlength{\Cscr}{\value{CEht}\Ctenthex}%
\addtolength{\Cscr}{-1.0ex}\protect\raisebox{0ex}[\value{CEht}%
\Ctenthex][\Cscr]{}\protect\end{array}\protect\right.\protect
\label{eq:Physical-ConvRenormConstantA}\protect\end{eqnarray}%
\setcounter{CEht}{10}where the symbol ``\raisebox{0ex}[1ex][0ex]%
{$\protect\displaystyle\:\!\!{\protect\it\Lambda\!\:}$}'' undern%
eath the integral reminds us that we are integrating over a ``bo%
x'' of width \raisebox{0ex}[1ex][0ex]{$\protect\displaystyle\:\!%
\!{\protect\it\Lambda\!\:}$} in momentum space, namely, the firs%
t Brillouin zone (regardless of how we choose its boundaries). T%
he result (\protect\ref{eq:Physical-ConvRenormConstantA}) is the 
natural definition of the convolution of two continuous fields, 
as it corresponds to an ``averaging'' process, with the factor o%
f \raisebox{0ex}[1ex][0ex]{$\protect\displaystyle1/\:\!\!{%
\protect\it\Lambda\!\:}$} balancing the integration over a box o%
f width \raisebox{0ex}[1ex][0ex]{$\protect\displaystyle\:\!\!{%
\protect\it\Lambda\!\:}$}.\par We can now investigate how the in%
verse Fourier transform (\protect\ref{eq:Fourier-InverseFourier}%
) relates position space and momentum space in the case of const%
ant $a$ as \raisebox{0ex}[1ex][0ex]{$\protect\displaystyle N%
\rightarrow\infty$}. Using the renormalised function (\protect
\ref{eq:Physical-RenormalisedConstantA}) via \mbox{Eq.~\raisebox
{0ex}[1ex][0ex]{$\protect\displaystyle\!$}}(\protect\ref{eq:Phys%
ical-InvertedRenormalisedConstantA}), \mbox{Eq.~\raisebox{0ex}[1%
ex][0ex]{$\protect\displaystyle\!$}}(\protect\ref{eq:Fourier-Inv%
erseFourier}) gives \setcounter{Ceqindent}{0}\protect\begin{eqna%
rray}\protect\left.\protect\begin{array}{rcl}\protect
\displaystyle\hspace{-1.3ex}&\protect\displaystyle f_{\:\!\!n}=%
\lim_{N\rightarrow\infty}\mbox{$\protect\displaystyle\protect
\frac{1}{\delta^{\:\!\!}p\:\!\sqrt N}$}\sum_b\:\!\!\delta^{\:\!%
\!}p\,e^{2\pi ibn/N\!}\mbox{$\protect\displaystyle\protect\frac{%
\,\overline{\!F}_{\!b}}{\sqrt N}$}=\lim_{N\rightarrow\infty}\mbox
{$\protect\displaystyle\protect\frac{a}{2\pi}$}^{\!}\sum_b\:\!\!%
\delta^{\:\!\!}p\,e^{2\pi ibn/N\:\!\!}\,\overline{\!F}_{\!b},%
\setlength{\Cscr}{\value{CEht}\Ctenthex}\addtolength{\Cscr}{-1.0%
ex}\protect\raisebox{0ex}[\value{CEht}\Ctenthex][\Cscr]{}\protect
\end{array}\protect\right.\protect\label{eq:Physical-InvFourFirs%
tConstantA}\protect\end{eqnarray}\setcounter{CEht}{10}so that 
\setcounter{Ceqindent}{0}\protect\begin{eqnarray}\protect\left.%
\protect\begin{array}{rcl}\protect\displaystyle\hspace{-1.3ex}&%
\protect\displaystyle f_{\:\!\!n}=\mbox{$\protect\displaystyle
\protect\frac{1}{{\protect\it\Lambda\!\:}^{\:\!}}$}^{\!}\hspace{%
-0.5mm}{\protect\mbox{}}\hspace{-0.1mm}\makebox[0ex]{\raisebox{-%
5.4mm}{\hspace{-0ex}\hspace{1mm}\makebox[0ex]{\scriptsize{%
\protect\mbox{}}${\protect\it\Lambda\!\:}$}}\hspace{0ex}\hspace{%
-1.5mm}}\protect\int{\protect\mbox{}}\hspace{-0.5mm}\hspace{-0.6%
mm}dp\,e^{ipx_{\:\!\!n\:\!\!}}\,\overline{\!F}(p).\setlength{%
\Cscr}{\value{CEht}\Ctenthex}\addtolength{\Cscr}{-1.0ex}\protect
\raisebox{0ex}[\value{CEht}\Ctenthex][\Cscr]{}\protect\end{array%
}\protect\right.\protect\label{eq:Physical-InverseFourierConstan%
tA}\protect\end{eqnarray}\setcounter{CEht}{10}Thus, by using the 
renormalised function \raisebox{0ex}[1ex][0ex]{$\protect
\displaystyle\,\overline{\!F}(p)$} in momentum space, we have au%
tomatically obtained a sensible result for the inverse Fourier t%
ransform. Likewise, \mbox{Eq.~\raisebox{0ex}[1ex][0ex]{$\protect
\displaystyle\!$}}(\protect\ref{eq:Fourier-Fourier}) gives 
\setcounter{Ceqindent}{0}\protect\begin{eqnarray}\protect\left.%
\protect\begin{array}{rcl}\protect\displaystyle\hspace{-1.3ex}&%
\protect\displaystyle\mbox{$\protect\displaystyle\protect\frac{%
\,\overline{\!F}_{\!b}}{\sqrt N}$}=\mbox{$\protect\displaystyle
\protect\frac{1}{\sqrt N}$}\!\:\!\!\sum_{n>-\:\!\!N\:\!\!/2}^{%
\leq N\:\!\!/2}\!\!e^{-2\pi ibn/N\!}f_{\:\!\!n},\setlength{\Cscr
}{\value{CEht}\Ctenthex}\addtolength{\Cscr}{-1.0ex}\protect
\raisebox{0ex}[\value{CEht}\Ctenthex][\Cscr]{}\protect\end{array%
}\protect\right.\protect\label{eq:Physical-FourFirstConstantA}%
\protect\end{eqnarray}\setcounter{CEht}{10}so that the limit 
\raisebox{0ex}[1ex][0ex]{$\protect\displaystyle N\rightarrow
\infty$} is similarly well-defined: \setcounter{Ceqindent}{0}%
\protect\begin{eqnarray}\protect\left.\protect\begin{array}{rcl}%
\protect\displaystyle\hspace{-1.3ex}&\protect\displaystyle\,%
\overline{\!F}(p)=\!\!\sum_{n=-\infty}^\infty\!\!e^{-ipx_{\:\!\!%
n}\!}f_{\:\!\!n}.\setlength{\Cscr}{\value{CEht}\Ctenthex}%
\addtolength{\Cscr}{-1.0ex}\protect\raisebox{0ex}[\value{CEht}%
\Ctenthex][\Cscr]{}\protect\end{array}\protect\right.\protect
\label{eq:Physical-FourierConstantA}\protect\end{eqnarray}%
\setcounter{CEht}{10}Finally, our use of renormalised functions 
in momentum space means that we need to refine the definition of 
the convolution operation in position space. Namely, if we form 
the product of two \mbox{}\protect\/{\protect\em renormalised%
\protect\/} functions \raisebox{0ex}[1ex][0ex]{$\protect
\displaystyle\,\overline{\!F}(p)$} and \raisebox{0ex}[1ex][0ex]{%
$\protect\displaystyle\,\overline{\!G}(p)$} in momentum space, t%
hen what is the corresponding operation between the functions 
\raisebox{0ex}[1ex][0ex]{$\protect\displaystyle f_{\:\!\!n}$} an%
d $g_n$ in position space? The simplest way to obtain the correc%
t result is to directly obtain the inverse Fourier transform of 
\raisebox{0ex}[1ex][0ex]{$\protect\displaystyle(\,\overline{\!F}%
\,\overline{\!G})_b$}. Using (\protect\ref{eq:Physical-InvFourFi%
rstConstantA}) without the introduction of \raisebox{0ex}[1ex][0%
ex]{$\protect\displaystyle\delta^{\:\!\!}p$}, together with (%
\protect\ref{eq:Physical-FourFirstConstantA}), we obtain 
\setcounter{Ceqindent}{0}\protect\begin{eqnarray}\mbox{$\protect
\displaystyle\protect\frac{1}{N}$}^{\!}\sum_be^{2\pi ibn/N\:\!\!%
}\,\overline{\!F}_{\!b\:\!}\,\overline{\!G}_b\hspace{-1.3ex}&%
\displaystyle=&\hspace{-1.3ex}\mbox{$\protect\displaystyle
\protect\frac{1}{N}$}^{\!}\sum_be^{+2\pi ibn/N\:\!\!}\sum_{n^{%
\prime\!}}e^{-2\pi ibn^{\prime\!}/N\!}f_{\:\!\!n^{\prime\!}}\sum
_{n^{\prime\:\!\!\prime\!}}e^{-2\pi ibn^{\prime\:\!\!\prime\!}/N%
\!}g_{n^{\prime\:\!\!\prime\!}}\protect\nonumber\setlength{\Cscr
}{\value{CEht}\Ctenthex}\addtolength{\Cscr}{-1.0ex}\protect
\raisebox{0ex}[\value{CEht}\Ctenthex][\Cscr]{}\\*[0ex]\protect
\displaystyle\hspace{-1.3ex}&\displaystyle=&\hspace{-1.3ex}\mbox
{$\protect\displaystyle\protect\frac{1}{N}$}^{\!}\sum_{n^{\prime
\!}}f_{\:\!\!n^{\prime\!}}\sum_{n^{\prime\:\!\!\prime\!}}g_{n^{%
\prime\:\!\!\prime\!}}\sum_be^{2\pi ib(n-n^{\prime\!}-n^{\prime
\:\!\!\prime\!}{}^{\:\!})/N}\protect\nonumber\setlength{\Cscr}{%
\value{CEht}\Ctenthex}\addtolength{\Cscr}{-1.0ex}\protect
\raisebox{0ex}[\value{CEht}\Ctenthex][\Cscr]{}\\*[0ex]\protect
\displaystyle\hspace{-1.3ex}&\displaystyle=&\hspace{-1.3ex}\mbox
{$\protect\displaystyle\protect\frac{1}{N}$}^{\!}\sum_{n^{\prime
\!}}f_{\:\!\!n^{\prime\!}}\sum_{n^{\prime\:\!\!\prime\!}}g_{n^{%
\prime\:\!\!\prime\!}}\,N^{\:\!}\delta_{n^{\prime\:\!\!\prime\!}%
\:\!\!,\,n-n^{\prime\!}}\protect\nonumber\setlength{\Cscr}{\value
{CEht}\Ctenthex}\addtolength{\Cscr}{-1.0ex}\protect\raisebox{0ex%
}[\value{CEht}\Ctenthex][\Cscr]{}\\*[0ex]\protect\displaystyle
\hspace{-1.3ex}&\displaystyle=&\hspace{-1.3ex}\sum_{n^{\prime\!}%
}f_{\:\!\!n^{\prime\!}}\,g_{n-n^{\prime\!}}.\protect\nonumber
\setlength{\Cscr}{\value{CEht}\Ctenthex}\addtolength{\Cscr}{-1.0%
ex}\protect\raisebox{0ex}[\value{CEht}\Ctenthex][\Cscr]{}\protect
\end{eqnarray}\setcounter{CEht}{10}Thus, taking the limit 
\raisebox{0ex}[1ex][0ex]{$\protect\displaystyle N\rightarrow
\infty$}, and defining \setcounter{Ceqindent}{0}\protect\begin{e%
qnarray}\hspace{-1.3ex}&\displaystyle{{\cal F}}^{-1\:\!\!}\{(\,%
\overline{\!F}\,\overline{\!G})^{\:\!\!}(p)\}\equiv(f\!\ast\!g)_%
n,\protect\nonumber\setlength{\Cscr}{\value{CEht}\Ctenthex}%
\addtolength{\Cscr}{-1.0ex}\protect\raisebox{0ex}[\value{CEht}%
\Ctenthex][\Cscr]{}\protect\end{eqnarray}\setcounter{CEht}{10}we 
find that the position-space convolution operation is also compl%
etely well-defined: \setcounter{Ceqindent}{0}\protect\begin{eqna%
rray}\protect\left.\protect\begin{array}{rcl}\protect
\displaystyle\hspace{-1.3ex}&\protect\displaystyle(f\!\ast\!g)_n%
=\!\!\sum_{n'=-\infty}^\infty\!\!f_{\:\!\!n^{\prime\!}}\,g_{n-n^%
{\prime\!}}.\setlength{\Cscr}{\value{CEht}\Ctenthex}\addtolength
{\Cscr}{-1.0ex}\protect\raisebox{0ex}[\value{CEht}\Ctenthex][%
\Cscr]{}\protect\end{array}\protect\right.\protect\label{eq:Phys%
ical-PositionConvolutionConstantA}\protect\end{eqnarray}%
\setcounter{CEht}{10}\par So far we have examined only the case 
in which the \raisebox{0ex}[1ex][0ex]{$\protect\displaystyle n%
\rightarrow\infty$} limit has been taken with the lattice spacin%
g kept physically constant, namely, \mbox{Eq.~\raisebox{0ex}[1ex%
][0ex]{$\protect\displaystyle\!$}}(\protect\ref{eq:Physical-Cons%
tantA}), which gives us an infinite lattice in position space, a%
nd a continuum limited to the first Brillouin zone in momentum s%
pace. The second simple prescription we may make for the 
\raisebox{0ex}[1ex][0ex]{$\protect\displaystyle N\rightarrow
\infty$} limit is to assume that the \mbox{}\protect\/{\protect
\em length\protect\/} of the position-space lattice, $L$, remain%
s constant for all $N$: \setcounter{Ceqindent}{0}\protect\begin{%
eqnarray}\protect\left.\protect\begin{array}{rcl}\protect
\displaystyle\hspace{-1.3ex}&\protect\displaystyle L(N)=L=\mbox{%
constant}.\setlength{\Cscr}{\value{CEht}\Ctenthex}\addtolength{%
\Cscr}{-1.0ex}\protect\raisebox{0ex}[\value{CEht}\Ctenthex][\Cscr
]{}\protect\end{array}\protect\right.\protect\label{eq:Physical-%
ConstantL}\protect\end{eqnarray}\setcounter{CEht}{10}{}From\ (%
\protect\ref{eq:Physical-Length}), this implies that $a(N)$ is i%
nversely proportional to $N$: \setcounter{Ceqindent}{0}\protect
\begin{eqnarray}\protect\left.\protect\begin{array}{rcl}\protect
\displaystyle\hspace{-1.3ex}&\protect\displaystyle a(N)=\mbox{$%
\protect\displaystyle\protect\frac{L}{N}$},\hspace{6ex}\mbox{%
\mbox{}\protect\/{\protect\em i.e.\protect\/}, }\hspace{2ex}a(N)%
\propto\mbox{$\protect\displaystyle\protect\frac{1}{N}$}.%
\setlength{\Cscr}{\value{CEht}\Ctenthex}\addtolength{\Cscr}{-1.0%
ex}\protect\raisebox{0ex}[\value{CEht}\Ctenthex][\Cscr]{}\protect
\end{array}\protect\right.\protect\label{eq:Physical-AConstantL}%
\protect\end{eqnarray}\setcounter{CEht}{10}\mbox{Eq.~\raisebox{0%
ex}[1ex][0ex]{$\protect\displaystyle\!$}}(\protect\ref{eq:Physic%
al-DefineLambda}) now tells is that it is the ``length'' of the 
lattice in \mbox{}\protect\/{\protect\em momentum\protect\/} spa%
ce, ${\protect\it\Lambda\!\:}$, that is proportional to $N$: 
\setcounter{Ceqindent}{0}\protect\begin{eqnarray}\protect\left.%
\protect\begin{array}{rcl}\protect\displaystyle\hspace{-1.3ex}&%
\protect\displaystyle{\protect\it\Lambda\!\:}(N)=\mbox{$\protect
\displaystyle\protect\frac{2\pi^{\:\!\!}N}{L}$},\hspace{6ex}\mbox
{\mbox{}\protect\/{\protect\em i.e.\protect\/}, }\hspace{2ex}{%
\protect\it\Lambda\!\:}(N)\propto N,\setlength{\Cscr}{\value{CEh%
t}\Ctenthex}\addtolength{\Cscr}{-1.0ex}\protect\raisebox{0ex}[%
\value{CEht}\Ctenthex][\Cscr]{}\protect\end{array}\protect\right
.\protect\label{eq:Physical-LambdaConstantL}\protect\end{eqnarra%
y}\setcounter{CEht}{10}whereas it is the ``spacing'' in momentum 
space that is now a constant: \setcounter{Ceqindent}{0}\protect
\begin{eqnarray}\protect\left.\protect\begin{array}{rcl}\protect
\displaystyle\hspace{-1.3ex}&\protect\displaystyle\delta^{\:\!\!%
}p=\mbox{$\protect\displaystyle\protect\frac{2\pi}{L}$}.%
\setlength{\Cscr}{\value{CEht}\Ctenthex}\addtolength{\Cscr}{-1.0%
ex}\protect\raisebox{0ex}[\value{CEht}\Ctenthex][\Cscr]{}\protect
\end{array}\protect\right.\protect\label{eq:Physical-DeltaPConst%
antL}\protect\end{eqnarray}\setcounter{CEht}{10}In other words, 
we have the complete conjugate of the previous scenario: we have 
a continuum in \mbox{}\protect\/{\protect\em position\protect\/} 
space, contained within a ``box'' of width $L$, which correspond%
s to an infinite lattice of states in \mbox{}\protect\/{\protect
\em momentum\protect\/} space. Because of the symmetry of the (f%
inite lattice) Fourier transform, we automatically know that, in 
this case, we will have need to renormalise functions in \mbox{}%
\protect\/{\protect\em position\protect\/} space: \setcounter{Ce%
qindent}{0}\protect\begin{eqnarray}\protect\left.\protect\begin{%
array}{rcl}\protect\displaystyle\hspace{-1.3ex}&\protect
\displaystyle\:\!\overline{\:\!\!h\:\!\!}\:\!(x)\equiv h(x)\sqrt
N,\setlength{\Cscr}{\value{CEht}\Ctenthex}\addtolength{\Cscr}{-1%
.0ex}\protect\raisebox{0ex}[\value{CEht}\Ctenthex][\Cscr]{}%
\protect\end{array}\protect\right.\protect\label{eq:Physical-Ren%
ormalisedConstantL}\protect\end{eqnarray}\setcounter{CEht}{10}so 
that \setcounter{Ceqindent}{0}\protect\begin{eqnarray}\protect
\left.\protect\begin{array}{rcl}\protect\displaystyle\hspace{-1.%
3ex}&\protect\displaystyle h(x)=\mbox{$\protect\displaystyle
\protect\frac{\:\!\overline{\:\!\!h\:\!\!}\:\!(x)}{\sqrt N}$}.%
\setlength{\Cscr}{\value{CEht}\Ctenthex}\addtolength{\Cscr}{-1.0%
ex}\protect\raisebox{0ex}[\value{CEht}\Ctenthex][\Cscr]{}\protect
\end{array}\protect\right.\protect\label{eq:Physical-InvertedRen%
ormalisedConstantL}\protect\end{eqnarray}\setcounter{CEht}{10}No%
w, using the fact that the spacing between lattice sites in posi%
tion space is just \setcounter{Ceqindent}{0}\protect\begin{eqnar%
ray}\hspace{-1.3ex}&\displaystyle\delta x\equiv a,\protect
\nonumber\setlength{\Cscr}{\value{CEht}\Ctenthex}\addtolength{%
\Cscr}{-1.0ex}\protect\raisebox{0ex}[\value{CEht}\Ctenthex][\Cscr
]{}\protect\end{eqnarray}\setcounter{CEht}{10}the convolution op%
eration in position space, (\protect\ref{eq:Fourier-ConvolutionO%
riginal}), becomes \setcounter{Ceqindent}{0}\protect\begin{eqnar%
ray}\hspace{-1.3ex}&\displaystyle\,\overline{\!\:\!(f\!\ast\!g)%
\:\!\!}_{\:\!n}=\mbox{$\protect\displaystyle\protect\frac{1}{N^{%
\!}a}$}\sum_{n'}{}^{\!}a\,\,\overline{\!f}_{\:\!\!\!n^{\prime\!}%
\;}\:\!\overline{\:\!\!g}_{n-n^{\prime\!}},\protect\nonumber
\setlength{\Cscr}{\value{CEht}\Ctenthex}\addtolength{\Cscr}{-1.0%
ex}\protect\raisebox{0ex}[\value{CEht}\Ctenthex][\Cscr]{}\protect
\end{eqnarray}\setcounter{CEht}{10}which in the limit \raisebox{%
0ex}[1ex][0ex]{$\protect\displaystyle N\rightarrow\infty$} gives 
\setcounter{Ceqindent}{0}\protect\begin{eqnarray}\protect\left.%
\protect\begin{array}{rcl}\protect\displaystyle\hspace{-1.3ex}&%
\protect\displaystyle\,\overline{\!\:\!(f\!\ast\!g)}(x)=\mbox{$%
\protect\displaystyle\protect\frac{1}{L}$}\hspace{-0.5mm}{%
\protect\mbox{}}\hspace{-0.1mm}\makebox[0ex]{\raisebox{-5.4mm}{%
\hspace{-0ex}\hspace{1mm}\makebox[0ex]{\scriptsize{\protect\mbox
{}}$L$}}\hspace{0ex}\hspace{-1.5mm}}\protect\int{\protect\mbox{}%
}\hspace{-0.5mm}\hspace{-0.6mm}dx\,\,\overline{\!f}^{\:\!\!}(x')%
^{\,}\:\!\overline{\:\!\!g}(x\!-\!x'),\setlength{\Cscr}{\value{C%
Eht}\Ctenthex}\addtolength{\Cscr}{-1.0ex}\protect\raisebox{0ex}[%
\value{CEht}\Ctenthex][\Cscr]{}\protect\end{array}\protect\right
.\protect\label{eq:Physical-ConvRenormConstantL}\protect\end{eqn%
array}\setcounter{CEht}{10}where again the factor of $1/L$ natur%
ally balances the integration over a box of width $L$; \mbox{}%
\protect\/{\protect\em i.e.\protect\/}, we are computing an aver%
age over the width of the box.\par Clearly, when we write the re%
sults in this way, the symmetry between position space and momen%
tum space is manifest. This allows us to write down the constant%
-$L$ analogues of (\protect\ref{eq:Physical-InverseFourierConsta%
ntA}), (\protect\ref{eq:Physical-FourierConstantA})\ and (%
\protect\ref{eq:Physical-PositionConvolutionConstantA}) directly%
: \setcounter{Ceqindent}{0}\protect\begin{eqnarray}\hspace{-1.3e%
x}&\displaystyle F_b=\mbox{$\protect\displaystyle\protect\frac{1%
}{L}$}\hspace{-0.5mm}{\protect\mbox{}}\hspace{-0.1mm}\makebox[0e%
x]{\raisebox{-5.4mm}{\hspace{-0ex}\hspace{1mm}\makebox[0ex]{%
\scriptsize{\protect\mbox{}}$L$}}\hspace{0ex}\hspace{-1.5mm}}%
\protect\int{\protect\mbox{}}\hspace{-0.5mm}\hspace{-0.6mm}dx\,e%
^{-ip_bx}\,\overline{\!f}^{\:\!\!}(x),\protect\label{eq:Physical%
-FourierConstantL}\setlength{\Cscr}{\value{CEht}\Ctenthex}%
\addtolength{\Cscr}{-1.0ex}\protect\raisebox{0ex}[\value{CEht}%
\Ctenthex][\Cscr]{}\\*[0ex]\protect\displaystyle\hspace{-1.3ex}&%
\displaystyle\,\overline{\!f}(x)=\!\!\sum_{b=-\infty}^\infty\!\!%
e^{-ip_bx}F_b,\protect\label{eq:Physical-InverseFourierConstantL%
}\setlength{\Cscr}{\value{CEht}\Ctenthex}\addtolength{\Cscr}{-1.%
0ex}\protect\raisebox{0ex}[\value{CEht}\Ctenthex][\Cscr]{}\\*[0e%
x]\protect\displaystyle\hspace{-1.3ex}&\displaystyle(F\!\ast\!G)%
_b=\!\!\sum_{b'=-\infty}^\infty\!\!\!F_{\:\!\!b^{\prime\!}}\,G_{%
b-b^{\prime\!}},\protect\label{eq:Physical-MomentumConvolutionCo%
nstantL}\setlength{\Cscr}{\value{CEht}\Ctenthex}\addtolength{%
\Cscr}{-1.0ex}\protect\raisebox{0ex}[\value{CEht}\Ctenthex][\Cscr
]{}\protect\end{eqnarray}\setcounter{CEht}{10}which can be confi%
rmed by straightforward calculations.\par Finally, we consider t%
he case in which we want both position \mbox{}\protect\/{\protect
\em and\protect\/} momentum space to become continua in the limi%
t \raisebox{0ex}[1ex][0ex]{$\protect\displaystyle N\rightarrow
\infty$}. Again, there are many ways in which this can be done, 
but the simplest to analyse is the symmetrical prescription 
\setcounter{Ceqindent}{0}\protect\begin{eqnarray}\protect\left.%
\protect\begin{array}{rcl}\protect\displaystyle\hspace{-1.3ex}&%
\protect\displaystyle a(N)\propto\mbox{$\protect\displaystyle
\protect\frac{1}{\sqrt N}$},\setlength{\Cscr}{\value{CEht}%
\Ctenthex}\addtolength{\Cscr}{-1.0ex}\protect\raisebox{0ex}[%
\value{CEht}\Ctenthex][\Cscr]{}\protect\end{array}\protect\right
.\protect\label{eq:Physical-SymmPrescription}\protect\end{eqnarr%
ay}\setcounter{CEht}{10}so that the size of the ``box'' in both 
position space and momentum space is proportional to $\sqrt N$: 
\setcounter{Ceqindent}{0}\protect\begin{eqnarray}\protect\left.%
\protect\begin{array}{rcl}\protect\displaystyle\hspace{-1.3ex}&%
\protect\displaystyle L(N)\propto\sqrt N,\hspace{15ex}{\protect
\it\Lambda\!\:}^{\:\!\!}(N)\propto\sqrt N,\setlength{\Cscr}{%
\value{CEht}\Ctenthex}\addtolength{\Cscr}{-1.0ex}\protect
\raisebox{0ex}[\value{CEht}\Ctenthex][\Cscr]{}\protect\end{array%
}\protect\right.\protect\label{eq:Physical-SymmBoxSizes}\protect
\end{eqnarray}\setcounter{CEht}{10}and the spacing between adjac%
ent lattice sites in both position and momentum space is inverse%
ly proportional to $\sqrt N$: \setcounter{Ceqindent}{0}\protect
\begin{eqnarray}\protect\left.\protect\begin{array}{rcl}\protect
\displaystyle\hspace{-1.3ex}&\protect\displaystyle\delta^{\:\!\!%
}x(N)\propto\mbox{$\protect\displaystyle\protect\frac{1}{\sqrt N%
}$},\hspace{15ex}\delta^{\:\!\!}p(N)\propto\mbox{$\protect
\displaystyle\protect\frac{1}{\sqrt N}$}.\setlength{\Cscr}{\value
{CEht}\Ctenthex}\addtolength{\Cscr}{-1.0ex}\protect\raisebox{0ex%
}[\value{CEht}\Ctenthex][\Cscr]{}\protect\end{array}\protect
\right.\protect\label{eq:Physical-SymmSpacings}\protect\end{eqna%
rray}\setcounter{CEht}{10}The prescription (\protect\ref{eq:Phys%
ical-SymmPrescription}) requires us to fix the constant of propo%
rtionality---say, \setcounter{Ceqindent}{0}\protect\begin{eqnarr%
ay}\hspace{-1.3ex}&\displaystyle a(N)=\mbox{$\protect
\displaystyle\protect\frac{a_0}{\sqrt N}$},\protect\nonumber
\setlength{\Cscr}{\value{CEht}\Ctenthex}\addtolength{\Cscr}{-1.0%
ex}\protect\raisebox{0ex}[\value{CEht}\Ctenthex][\Cscr]{}\protect
\end{eqnarray}\setcounter{CEht}{10}where $a_0$ is some constant 
with the dimensions of length. However, it is practically conven%
ient for us to choose our system of units so that \raisebox{0ex}%
[1ex][0ex]{$\protect\displaystyle a_0\equiv1$}, so that (\protect
\ref{eq:Physical-SymmBoxSizes}) become \addtocounter{equation}{-%
1}\renewcommand\theequation{\protect\ref{eq:Physical-SymmBoxSize%
s}$'$}\setcounter{Ceqindent}{0}\protect\begin{eqnarray}\protect
\left.\protect\begin{array}{rcl}\protect\displaystyle\hspace{-1.%
3ex}&\protect\displaystyle L(N)=\sqrt N,\hspace{15ex}{\protect\it
\Lambda\!\:}^{\:\!\!}(N)=2\pi\sqrt N,\setlength{\Cscr}{\value{CE%
ht}\Ctenthex}\addtolength{\Cscr}{-1.0ex}\protect\raisebox{0ex}[%
\value{CEht}\Ctenthex][\Cscr]{}\protect\end{array}\protect\right
.\protect\end{eqnarray}\renewcommand\theequation{\arabic{equatio%
n}}\setcounter{CEht}{10}and (\protect\ref{eq:Physical-SymmSpacin%
gs}) become \addtocounter{equation}{-1}\renewcommand\theequation
{\protect\ref{eq:Physical-SymmSpacings}$'$}\setcounter{Ceqindent%
}{0}\protect\begin{eqnarray}\protect\left.\protect\begin{array}{%
rcl}\protect\displaystyle\hspace{-1.3ex}&\protect\displaystyle
\delta^{\:\!\!}x(N)=\mbox{$\protect\displaystyle\protect\frac{1}%
{\sqrt N}$},\hspace{15ex}\delta^{\:\!\!}p(N)=\mbox{$\protect
\displaystyle\protect\frac{2\pi}{\sqrt N}$}.\setlength{\Cscr}{%
\value{CEht}\Ctenthex}\addtolength{\Cscr}{-1.0ex}\protect
\raisebox{0ex}[\value{CEht}\Ctenthex][\Cscr]{}\protect\end{array%
}\protect\right.\protect\end{eqnarray}\renewcommand\theequation{%
\arabic{equation}}\setcounter{CEht}{10}If this is done, the 
\raisebox{0ex}[1ex][0ex]{$\protect\displaystyle N\rightarrow
\infty$} results can be written down directly from (\protect\ref
{eq:Fourier-Fourier}), (\protect\ref{eq:Fourier-InverseFourier})%
, (\protect\ref{eq:Fourier-ConvolutionFourier})\ and (\protect
\ref{eq:Fourier-ConvolutionOriginal}): \setcounter{Ceqindent}{0}%
\protect\begin{eqnarray}\hspace{-1.3ex}&\displaystyle F(p)=\!%
\hspace{-0.5mm}{\protect\mbox{}}\hspace{-0.1mm}\protect\int_{\:%
\!\!-\infty}^{\infty}{\protect\mbox{}}\hspace{-0.5mm}\hspace{-0.%
6mm}dx\,e^{-ipx}f^{\:\!\!}(x),\protect\label{eq:Physical-Fourier%
}\setlength{\Cscr}{\value{CEht}\Ctenthex}\addtolength{\Cscr}{-1.%
0ex}\protect\raisebox{0ex}[\value{CEht}\Ctenthex][\Cscr]{}\\*[0e%
x]\protect\displaystyle\hspace{-1.3ex}&\displaystyle f^{\:\!\!}(%
x)=\mbox{$\protect\displaystyle\protect\frac{1}{2\pi}$}\hspace{-%
0.5mm}{\protect\mbox{}}\hspace{-0.1mm}\protect\int_{\:\!\!-\infty
}^{\infty}{\protect\mbox{}}\hspace{-0.5mm}\hspace{-0.6mm}dp\,e^{%
ipx}F(p),\protect\label{eq:Physical-InverseFourier}\setlength{%
\Cscr}{\value{CEht}\Ctenthex}\addtolength{\Cscr}{-1.0ex}\protect
\raisebox{0ex}[\value{CEht}\Ctenthex][\Cscr]{}\\*[0ex]\protect
\displaystyle\hspace{-1.3ex}&\displaystyle(F^{\:\!\!}\!\ast\!G)(%
p)=\mbox{$\protect\displaystyle\protect\frac{1}{2\pi}$}\hspace{-%
0.5mm}{\protect\mbox{}}\hspace{-0.1mm}\protect\int_{\:\!\!-\infty
}^{\infty}{\protect\mbox{}}\hspace{-0.5mm}\hspace{-0.6mm}dp'\:\!%
 F(p')^{\,}G(p\!-\!p'),\protect\label{eq:Physical-ConvolutionMom%
entum}\setlength{\Cscr}{\value{CEht}\Ctenthex}\addtolength{\Cscr
}{-1.0ex}\protect\raisebox{0ex}[\value{CEht}\Ctenthex][\Cscr]{}%
\\*[0ex]\protect\displaystyle\hspace{-1.3ex}&\displaystyle(f\!%
\ast\!g)(x)=\!\hspace{-0.5mm}{\protect\mbox{}}\hspace{-0.1mm}%
\protect\int_{\:\!\!-\infty}^{\infty}{\protect\mbox{}}\hspace{-0%
.5mm}\hspace{-0.6mm}dx'\:\!f^{\:\!\!}(x')^{\:\!}g(x\!-\!x'),%
\protect\label{eq:Physical-ConvolutionPosition}\setlength{\Cscr}%
{\value{CEht}\Ctenthex}\addtolength{\Cscr}{-1.0ex}\protect
\raisebox{0ex}[\value{CEht}\Ctenthex][\Cscr]{}\protect\end{eqnar%
ray}\setcounter{CEht}{10}with no need to ``renormalise'' functio%
ns in either position space or momentum space. It is an accident 
of pedagogical history---that should probably be rectified---tha%
t the continuum Fourier transform of \mbox{Eqs.~\raisebox{0ex}[1%
ex][0ex]{$\protect\displaystyle\!$}}(\protect\ref{eq:Physical-Fo%
urier}), (\protect\ref{eq:Physical-InverseFourier}), (\protect
\ref{eq:Physical-ConvolutionMomentum})\ and (\protect\ref{eq:Phy%
sical-ConvolutionPosition}) is often taught to undergraduate stu%
dents without providing a rigorous justification on the basis of 
it being the symmetrical \raisebox{0ex}[1ex][0ex]{$\protect
\displaystyle N\rightarrow\infty$} limit (\protect\ref{eq:Physic%
al-SymmPrescription}) of the discrete Fourier transform. However%
, with the growing importance of lattice field theory, perhaps t%
his prejudice will slowly be overcome.\par For practical lattice 
calculations, of course, the dependence of $a(N)$ on $N$ does no%
t fall neatly into any of the three simple categories listed abo%
ve; indeed, one of the end-products of a lattice calculation is 
an estimate of $a$ based on physically identifiable results. The 
\raisebox{0ex}[1ex][0ex]{$\protect\displaystyle a\rightarrow0$} 
and \raisebox{0ex}[1ex][0ex]{$\protect\displaystyle L\rightarrow
\infty$} limits are still extrapolated, but they are not both ti%
ed symmetrically to $N$ in the way described by (\protect\ref{eq%
:Physical-SymmPrescription}); rather, both $N$ and the gauge cou%
pling $\beta$ are varied from calculation to calculation. This d%
oes not at all invalidate the connection made above between the 
lattice and continuum descriptions, but merely represents a more 
complicated way of taking the double limit than the simple presc%
ription (\protect\ref{eq:Physical-SymmPrescription}).\par
\refstepcounter{section}\vspace{1.5\baselineskip}\par{\centering
\bf\thesection. The problem for lattice field theory\\*[0.5%
\baselineskip]}\protect\indent\label{sect:Problem}In \mbox{Sec.~%
$\:\!\!$}\protect\ref{sect:Fourier} we looked at the relationshi%
p between a discrete lattice and its Fourier transform, from a p%
urely mathematical point of view. In \mbox{Sec.~$\:\!\!$}\protect
\ref{sect:Physical} we began to add some physical ``meat'' to th%
e description, in terms of identifying position space and moment%
um space, but the results were still essentially mathematical in 
nature. In this section we will look at where the properties of 
 Fourier theory become problematical for lattice field theory ca%
lculations in particular.\par The crux of the problem is the con%
volution operation. Even though we have set up our physical syst%
em to reside within a ``box''\mbox{$\!$}, in both position space 
and momentum space, the convolution operation goes outside this 
box and ``reaches around the other side'' of the lattice. Now, a%
s noted above, this is not a problem in position space, because 
we are using the fiction of position space being wrapped into a 
ring (in higher dimensions, of spacetime being wrapped into an $%
n$-torus) in order to avoid having rigid boundaries that would d%
estroy translation invariance and distort the calculations being 
performed inside the box. And, indeed, we find that for even the 
most elementary applications of a discrete space, we do wish to 
perform convolutions in position space; for example, the spatial 
derivative operator is really just a convolution, which correspo%
nds to multiplying the function in question by $ip$ in momentum 
space.\par It is with convolutions in momentum space that we hav%
e trouble. This corresponds to multiplying two functions in posi%
tion space, which is something that is only done for nonlinear a%
pplications---like quantum field theory. Now, we must expect tha%
t conventional Fourier transform wisdom applicable for linear ap%
plications \mbox{}\protect\/{\protect\em should\protect\/} need 
to be augmented when we delve into the nonlinear domain; and ind%
eed this is the case. When we multiply two fields, the high-mome%
ntum components of each field combine to ``alias'' (or ``masqera%
de'') as low-momentum components in the result. A simple example 
will suffice to demonstrate the problem. Imagine we have a field 
in the highest possible momentum eigenstate in the positive-$x$ 
direction, namely, \raisebox{0ex}[1ex][0ex]{$\protect
\displaystyle p=+\pi/a$}. Its spatial dependence will then be of 
the form \raisebox{0ex}[1ex][0ex]{$\protect\displaystyle e^{i\pi
x/a}$}\mbox{$\!$}. Imagine that we now have \mbox{}\protect\/{%
\protect\em another\protect\/} field, also in the highest possib%
le momentum eigenstate in the positive-$x$ direction. When we mu%
ltiply the two fields in position space, the product then has a 
spatial dependence of simply \raisebox{0ex}[1ex][0ex]{$\protect
\displaystyle e^{2i\pi x/a}$}\mbox{$\!$}, which corresponds to a 
momentum value of $+2\pi/a$. Indeed, in general, when we multipl%
y in position space two fields in momentum eigenstates of 
\raisebox{0ex}[1ex][0ex]{$\protect\displaystyle p_{1\!}$} and $p%
_2$, the product is in a momentum eigenstate of \raisebox{0ex}[1%
ex][0ex]{$\protect\displaystyle p=p_{1\!}+p_2$}. When one works 
it through, this simply represents the principle of conservation 
of momentum, when quantum field theory is interpreted in terms o%
f particles of definite momentum. However, consider the lattice 
representation of these two fields. For \raisebox{0ex}[1ex][0ex]%
{$\protect\displaystyle x_{n\:\!\!}=na$}, a spatial dependence o%
f \raisebox{0ex}[1ex][0ex]{$\protect\displaystyle e^{i\pi x/a\!}%
$} simply becomes \raisebox{0ex}[1ex][0ex]{$\protect\displaystyle
e^{i\pi n}\equiv(-1)^n$}\mbox{$\!$}, so that the field oscillate%
s \raisebox{0ex}[1ex][0ex]{$\protect\displaystyle+1,-1,+1,-1,%
\ldots,$} along the position-space lattice. If we multiply this 
by another field having the same spatial dependence, then we end 
up with \raisebox{0ex}[1ex][0ex]{$\protect\displaystyle+1,+1,+1,%
+1,\ldots.$} In other words, the product of the two fields has v%
anishing momentum! Where did the momentum of $2\pi/a$ disappear 
to? It was ``absorbed by the crystal in an Umklapp process''---e%
xcept that in quantum field theory we have no crystal. This zero%
-momentum eigenstate is a purely mathematical artefact.\par Anot%
her example might suffice to give a feel for the phenomenon of a%
liasing. Imagine that we have a field in a momentum eigenstate t%
hat is only $2/3$ of the maximum momentum that can be represente%
d on the lattice, namely, \raisebox{0ex}[1ex][0ex]{$\protect
\displaystyle p=2\pi/3a$}, so that it has a spatial dependence o%
f \raisebox{0ex}[1ex][0ex]{$\protect\displaystyle e^{2i\pi x/3a}%
$}\mbox{$\!$}. For \raisebox{0ex}[1ex][0ex]{$\protect
\displaystyle x_{n\:\!\!}=na$}, this is just \raisebox{0ex}[1ex]%
[0ex]{$\protect\displaystyle e^{2i\pi n/3}$}\mbox{$\!$}, which m%
eans that its position-space dependence is \setcounter{Ceqindent%
}{0}\protect\begin{eqnarray}\hspace{-1.3ex}&\displaystyle1,%
\hspace{3ex}-\mbox{$\protect\displaystyle\protect\frac{1}{2}$}+i%
\mbox{$\protect\displaystyle\protect\frac{\sqrt3}{2}$},\hspace{3%
ex}-\mbox{$\protect\displaystyle\protect\frac{1}{2}$}-i\mbox{$%
\protect\displaystyle\protect\frac{\sqrt3}{2}$},\hspace{3ex}1,%
\hspace{3ex}-\mbox{$\protect\displaystyle\protect\frac{1}{2}$}+i%
\mbox{$\protect\displaystyle\protect\frac{\sqrt3}{2}$},\hspace{3%
ex}-\mbox{$\protect\displaystyle\protect\frac{1}{2}$}-i\mbox{$%
\protect\displaystyle\protect\frac{\sqrt3}{2}$},\hspace{3ex}%
\ldots.\protect\nonumber\setlength{\Cscr}{\value{CEht}\Ctenthex}%
\addtolength{\Cscr}{-1.0ex}\protect\raisebox{0ex}[\value{CEht}%
\Ctenthex][\Cscr]{}\protect\end{eqnarray}\setcounter{CEht}{10}If 
we multiply this field by another field that is in the same mome%
ntum eigenstate of \raisebox{0ex}[1ex][0ex]{$\protect
\displaystyle p=2\pi/3a$}, then we again need to simply square e%
ach of these values: \setcounter{Ceqindent}{0}\protect\begin{eqn%
array}\hspace{-1.3ex}&\displaystyle1,\hspace{3ex}-\mbox{$\protect
\displaystyle\protect\frac{1}{2}$}-i\mbox{$\protect\displaystyle
\protect\frac{\sqrt3}{2}$},\hspace{3ex}-\mbox{$\protect
\displaystyle\protect\frac{1}{2}$}+i\mbox{$\protect\displaystyle
\protect\frac{\sqrt3}{2}$},\hspace{3ex}1,\hspace{3ex}-\mbox{$%
\protect\displaystyle\protect\frac{1}{2}$}-i\mbox{$\protect
\displaystyle\protect\frac{\sqrt3}{2}$},\hspace{3ex}-\mbox{$%
\protect\displaystyle\protect\frac{1}{2}$}+i\mbox{$\protect
\displaystyle\protect\frac{\sqrt3}{2}$},\hspace{3ex}\ldots.%
\protect\nonumber\setlength{\Cscr}{\value{CEht}\Ctenthex}%
\addtolength{\Cscr}{-1.0ex}\protect\raisebox{0ex}[\value{CEht}%
\Ctenthex][\Cscr]{}\protect\end{eqnarray}\setcounter{CEht}{10}We 
now recognise this to be simply \raisebox{0ex}[1ex][0ex]{$%
\protect\displaystyle e^{-2i\pi x/3a}$}; \mbox{}\protect\/{%
\protect\em i.e.\protect\/}, the product is in a momentum eigens%
tate of value \raisebox{0ex}[1ex][0ex]{$\protect\displaystyle p=%
-2\pi/3a$}. In the language of quantum field theory, two particl%
es of momentum \raisebox{0ex}[1ex][0ex]{$\protect\displaystyle p%
=+2\pi/3a$} have combined to create a particle of momentum 
\raisebox{0ex}[1ex][0ex]{$\protect\displaystyle p=-2\pi/3a$}!\par
Of course, the Umklapp analogy again tells us what is happening: 
the two particles combined to create a momentum eigenstate of 
\raisebox{0ex}[1ex][0ex]{$\protect\displaystyle p=+4\pi/3a$}, an%
d then the nonexistent ``crystal'' absorbed a momentum of 
\raisebox{0ex}[1ex][0ex]{$\protect\displaystyle2\pi/a$} to bring 
the result back into the first Brillouin zone.\par Let us now co%
nsider how we might repair this oversight.\par\refstepcounter{se%
ction}\vspace{1.5\baselineskip}\par{\centering\bf\thesection. A 
solution to the aliasing problem\\*[0.5\baselineskip]}\protect
\indent\label{sect:Solution}The engineer provides a simple metho%
dology for solving the aliasing problem: ensure that the fields 
are \mbox{}\protect\/{\protect\em filtered\protect\/} suitably, 
so that there is no ``cross-talk'' between aliased components an%
d genuine components in momentum space.\par So how do we carry o%
ut such a programme?\par The obvious ``knee-jerk'' reaction woul%
d be to filter every field with a low-pass filter, allowing thro%
ugh only momentum components \raisebox{0ex}[1ex][0ex]{$\protect
\displaystyle-\pi/2a<p\leq+\pi/2a$} that are no greater than 
\mbox{}\protect\/{\protect\em half\protect\/} the maximum possib%
le momentum (in absolute value) that can be represented on the l%
attice. When two such fields are multiplied together in position 
space, we can rest assured that the convolution process in momen%
tum space will \mbox{}\protect\/{\protect\em not\protect\/} ``bl%
eed out'' of the first Brillouin zone; the sum of any two moment%
a can just reach the Brillouin zone boundary, but it cannot exce%
ed it. This product of fields is then itself subjected to the sa%
me low-pass filtering, so that if we subsequently need to multip%
ly it by some \mbox{}\protect\/{\protect\em other\protect\/} (li%
kewise filtered) field, the result again is guaranteed to be con%
tained within the first Brillouin zone, without any aliasing art%
efacts.\par This process ensures that aliasing is removed comple%
tely, but it has cut down the available momentum modes too harsh%
ly---it is too conservative. Consider, instead, a low-pass filte%
r that allows through all momenta \raisebox{0ex}[1ex][0ex]{$%
\protect\displaystyle-2\pi/3a<p\leq+2\pi/3a$} that are no greate%
r than \mbox{}\protect\/{\protect\em two-thirds\protect\/} of th%
e maximum possible momentum value (in absolute value) that can b%
e represented on the lattice. If we calculate the position-space 
product of two fields filtered in this way, we know that the res%
ult will contain momentum values that rightly belong as high up 
as \raisebox{0ex}[1ex][0ex]{$\protect\displaystyle p=+4\pi/3a$} 
and as low as \raisebox{0ex}[1ex][0ex]{$\protect\displaystyle p=%
-4\pi/3a$}. The momentum components outside the first Brillouin 
zone will ``bleed out'' and be aliased in the first zone. Those 
momentum components that rightly belong in the range \raisebox{0%
ex}[1ex][0ex]{$\protect\displaystyle+\pi/a<p\leq+4\pi/3a$} will 
be ``aliased down'' by the subtraction of a ``crystal momentum'' 
of $2\pi/a$ into the range \raisebox{0ex}[1ex][0ex]{$\protect
\displaystyle-\pi/a<p\leq-2\pi/3a$}. Likewise, those components 
that rightly belong in the range \raisebox{0ex}[1ex][0ex]{$%
\protect\displaystyle-4\pi/3a<p\leq-\pi/a$} will be ``aliased up%
'' by addition of a ``crystal momentum'' of $2\pi/a$ into the ra%
nge \raisebox{0ex}[1ex][0ex]{$\protect\displaystyle+2\pi/3a<p\leq
+\pi/a$}. In other words, the momentum ranges \raisebox{0ex}[1ex%
][0ex]{$\protect\displaystyle-\pi/a<p\leq-2\pi/3a$} and \raisebox
{0ex}[1ex][0ex]{$\protect\displaystyle+2\pi/3a<p\leq+\pi/a$} are 
a mess: they are a combination of the genuine components of the 
product, polluted by the aliased components from the other end o%
f the spectrum.\par This does not seem to be much of an advance. 
However, consider what we now do: we \mbox{}\protect\/{\protect
\em filter the result\protect\/} with the same low-pass filter, 
so that it is ready for any further products that we may need to 
form in position space. This filters out the polluted momentum r%
anges completely; we are left with the components in the range 
\raisebox{0ex}[1ex][0ex]{$\protect\displaystyle-2\pi/3a<p\leq+2%
\pi/3a$} that represent the product with perfect fidelity! (By 
``perfect fidelity'' I do not mean that they represent the conti%
nuum exactly, because the lattice has provided a high-momentum r%
egulator that removes momentum modes outside this range from the 
outset; rather, we have a perfect representation of the \mbox{}%
\protect\/{\protect\em regulated\protect\/} theory, within this 
regulated momentum range.)\par In other words, the price we need 
to pay to remove aliasing from our lattice calculations is the c%
omplete elimination of one-third of the momentum modes in each d%
imension. For four-dimensional calculations, this means that we 
retain \raisebox{0ex}[1ex][0ex]{$\protect\displaystyle(2/3)^{4\!%
}=16/81$} of the modes originally present on a lattice of any gi%
ven dimensions, so that roughly four-fifths of them have been di%
scarded (there are lots of ``hypercorners'' in a hypercube!).\par
This reduction in momentum modes might seem to be wasteful: does%
n't this mean that we will need to spend five times as long perf%
orming the same lattice calculation with the same hardware? I mu%
st emphasise that nothing could be farther from the truth. The m%
omentum modes that we are eliminating are actually \mbox{}%
\protect\/{\protect\em distorting\protect\/} calculations that a%
re done without their elimination. Indeed, in recent years there 
has been a recognition of what has been called ``noise'' in the 
high-momentum components (short-distance structure) of lattice f%
ields, and attempts have been made to ``smooth'' the fields slig%
htly, to filter out some of this ``noise''\mbox{$\!$}. However, 
if this ``noise'' \mbox{}\protect\/{\protect\em was\protect\/} t%
ruly noise---uncorrelated statistical hiss---then it would not a%
ffect lattice calculations systematically at all; it would simpl%
y add to the inherent noise of the Monte Carlo process, and disa%
ppear in the mean. Aliasing, in contrast, does \mbox{}\protect\/%
{\protect\em not\protect\/} disappear on the average: it represe%
nts a form of ``wrap-around'' in momentum space, that can make f%
inite contributions systematically. Indeed, since loop diagrams 
in field theory inherently \mbox{}\protect\/{\protect\em diverge%
\protect\/} for high momentum, there is the potential for aliasi%
ng artefacts to completely swamp the genuine contributions to an%
y calculation.\par\refstepcounter{section}\vspace{1.5%
\baselineskip}\par{\centering\bf\thesection. Exact expressions f%
or the antialiasing operator\\*[0.5\baselineskip]}\protect\indent
\label{sect:Exact}In the previous section I argued that fields i%
n lattice field theory calculations should be passed regularly t%
hrough an ``antialiasing filter'' to eliminate all aliasing from 
the resulting calculations. The required filter is what the engi%
neers call a ``low-pass filter'': it allows through momentum com%
ponents in the range \raisebox{0ex}[1ex][0ex]{$\protect
\displaystyle-2\pi/3a<p\leq+2\pi/3a$} without distortion, and el%
iminates completely all components of higher momentum (in absolu%
te value).\par The position-space form of such an operator for a%
n infinite lattice can be derived by using the same method as wa%
s employed in [\ref{cit:Costella2002a}] for the SLAC derivative 
operator, and the corresponding finite-lattice expressions can b%
e derived by using the same method as was employed in [\ref{cit:%
Costella2002b}] to ``wrap'' the operator around the finite latti%
ce an infinite number of times. These are the tasks that we shal%
l carry out in this section.\par The ``trick'' of [\ref{cit:Cost%
ella2002a}] was to start with what we described in \mbox{Sec.~$%
\:\!\!$}\protect\ref{sect:Physical} as the ``constant $a$'' limi%
t of \raisebox{0ex}[1ex][0ex]{$\protect\displaystyle N\rightarrow
\infty$}, namely, we consider an infinite lattice in position sp%
ace with spacing $a$, which corresponds to a continuum in moment%
um space within the first Brillouin zone. Our antialiasing is th%
en simply described by \setcounter{Ceqindent}{0}\protect\begin{e%
qnarray}\protect\left.\protect\begin{array}{rcl}\protect
\displaystyle\hspace{-1.3ex}&\protect\displaystyle F(p)=\left\{%
\begin{array}{ll}1&\mbox{if $-2\pi/3a<p\leq+2\pi/3a$,}\\0&\mbox{%
otherwise.}\\\end{array}\right.\setlength{\Cscr}{\value{CEht}%
\Ctenthex}\addtolength{\Cscr}{-1.0ex}\protect\raisebox{0ex}[%
\value{CEht}\Ctenthex][\Cscr]{}\protect\end{array}\protect\right
.\protect\label{eq:Exact-Fp}\protect\end{eqnarray}\setcounter{CE%
ht}{10}The (``renormalised'') inverse Fourier transform (\protect
\ref{eq:Physical-InverseFourierConstantA}) can be performed in c%
losed form: \setcounter{Ceqindent}{0}\protect\begin{eqnarray}%
\hspace{-1.3ex}&\displaystyle f_{\:\!\!n}^{\infty\:\!\!}=\mbox{$%
\protect\displaystyle\protect\frac{a}{2\pi}$}\hspace{-0.5mm}{%
\protect\mbox{}}\hspace{-0.1mm}\protect\int_{-2\pi\:\!\!/3a}^{+2%
\pi\:\!\!/3a}{\protect\mbox{}}\hspace{-0.5mm}\hspace{-0.6mm}dp\,%
e^{ipx_{\:\!\!n}}=\mbox{$\protect\displaystyle\protect\frac{a}{2%
\pi ix_{\:\!\!n}}$}\setcounter{Cbscurr}{20}\setlength{\Cscr}{%
\value{Cbscurr}\Ctenthex}\addtolength{\Cscr}{-1.0ex}\protect
\raisebox{0ex}[\value{Cbscurr}\Ctenthex][\Cscr]{}\hspace{-0ex}{%
\protect\left\{\setlength{\Cscr}{\value{Cbscurr}\Ctenthex}%
\addtolength{\Cscr}{-1.0ex}\protect\raisebox{0ex}[\value{Cbscurr%
}\Ctenthex][\Cscr]{}\protect\right.}\hspace{-0.25ex}\setlength{%
\Cscr}{\value{Cbscurr}\Ctenthex}\addtolength{\Cscr}{-1.0ex}%
\protect\raisebox{0ex}[\value{Cbscurr}\Ctenthex][\Cscr]{}%
\setcounter{CbsD}{\value{CbsC}}\setcounter{CbsC}{\value{CbsB}}%
\setcounter{CbsB}{\value{CbsA}}\setcounter{CbsA}{\value{Cbscurr}%
}e^{+2\pi ix_{\:\!\!n\!}/3a}-e^{-2\pi ix_{\:\!\!n\!}/3a}%
\setlength{\Cscr}{\value{CbsA}\Ctenthex}\addtolength{\Cscr}{-1.0%
ex}\protect\raisebox{0ex}[\value{CbsA}\Ctenthex][\Cscr]{}\hspace
{-0.25ex}{\protect\left.\setlength{\Cscr}{\value{CbsA}\Ctenthex}%
\addtolength{\Cscr}{-1.0ex}\protect\raisebox{0ex}[\value{CbsA}%
\Ctenthex][\Cscr]{}\protect\right\}}\hspace{-0ex}\setlength{\Cscr
}{\value{CbsA}\Ctenthex}\addtolength{\Cscr}{-1.0ex}\protect
\raisebox{0ex}[\value{CbsA}\Ctenthex][\Cscr]{}\setcounter{CbsA}{%
\value{CbsB}}\setcounter{CbsB}{\value{CbsC}}\setcounter{CbsC}{%
\value{CbsD}}\setcounter{CbsD}{1},\protect\nonumber\setlength{%
\Cscr}{\value{CEht}\Ctenthex}\addtolength{\Cscr}{-1.0ex}\protect
\raisebox{0ex}[\value{CEht}\Ctenthex][\Cscr]{}\protect\end{eqnar%
ray}\setcounter{CEht}{10}which, from the definition of the sine 
function, and \raisebox{0ex}[1ex][0ex]{$\protect\displaystyle x_%
{n\!}\equiv na$}, yields \setcounter{Ceqindent}{0}\protect\begin
{eqnarray}\protect\left.\protect\begin{array}{rcl}\protect
\displaystyle\hspace{-1.3ex}&\protect\displaystyle f_{\:\!\!n}^{%
\infty\:\!\!}=\mbox{$\protect\displaystyle\protect\frac{a\,\sin(%
2\pi n/3)}{\pi n}$}.\setlength{\Cscr}{\value{CEht}\Ctenthex}%
\addtolength{\Cscr}{-1.0ex}\protect\raisebox{0ex}[\value{CEht}%
\Ctenthex][\Cscr]{}\protect\end{array}\protect\right.\protect
\label{eq:Exact-fnInfinite}\protect\end{eqnarray}\setcounter{CEh%
t}{10}(The superscript of ``$\infty$'' reminds us that we are wo%
rking with an infinite lattice at this stage.) Let us work in ``%
lattice units'' and set \raisebox{0ex}[1ex][0ex]{$\protect
\displaystyle a=1$}. To calculate \raisebox{0ex}[1ex][0ex]{$%
\protect\displaystyle f_{0\!}$} we need to take a limiting proce%
ss for \raisebox{0ex}[1ex][0ex]{$\protect\displaystyle n%
\rightarrow0$}, for which \raisebox{0ex}[1ex][0ex]{$\protect
\displaystyle\sin(2\pi n/3)\approx2\pi n/3$}, so that \raisebox{%
0ex}[1ex][0ex]{$\protect\displaystyle f_{0\!}=2/3$}. For all oth%
er \mbox{$n\equiv0$ (mod 3)}, \raisebox{0ex}[1ex][0ex]{$\protect
\displaystyle f_{\:\!\!n\!}$} is proportional to \raisebox{0ex}[%
1ex][0ex]{$\protect\displaystyle\sin(2\pi k)$} where \raisebox{0%
ex}[1ex][0ex]{$\protect\displaystyle k=n/3$}, and hence \raisebox
{0ex}[1ex][0ex]{$\protect\displaystyle f_{\:\!\!n\!}$} vanishes. 
 For \mbox{$n\equiv1$ (mod 3)}, we have \raisebox{0ex}[1ex][0ex]%
{$\protect\displaystyle\sin(2\pi/3\!+\!2\pi k)\equiv\sqrt3/2$}, 
so that \raisebox{0ex}[1ex][0ex]{$\protect\displaystyle f_{\:\!%
\!n\!}=\sqrt3/2\pi n$}. Likewise, for \mbox{$n\equiv2$ (mod 3)}, 
we have \raisebox{0ex}[1ex][0ex]{$\protect\displaystyle\sin(4\pi
/3\!+\!2\pi k)\equiv-\sqrt3/2$}, so that \raisebox{0ex}[1ex][0ex%
]{$\protect\displaystyle f_{\:\!\!n\!}=-\sqrt3/2\pi n$}. Thus, p%
utting these together, we have \setcounter{Ceqindent}{0}\protect
\begin{eqnarray}\protect\left.\protect\begin{array}{rcl}\protect
\displaystyle\hspace{-1.3ex}&\protect\displaystyle f_{\:\!\!n}^%
\infty=\left\{\begin{array}{ll}\hspace{3ex}\mbox{$\protect
\displaystyle\protect\frac{2}{3}$}\setlength{\Cscr}{36\Ctenthex}%
\addtolength{\Cscr}{-1.0ex}\protect\raisebox{0ex}[36\Ctenthex][%
\Cscr]{}&\mbox{if $n=0$,}\\+\mbox{$\protect\displaystyle\protect
\frac{\sqrt3}{2\pi n}$}\setlength{\Cscr}{40\Ctenthex}\addtolength
{\Cscr}{-1.0ex}\protect\raisebox{0ex}[40\Ctenthex][\Cscr]{}&\mbox
{if $n\equiv1$ (mod 3),}\\-\mbox{$\protect\displaystyle\protect
\frac{\sqrt3}{2\pi n}$}\setlength{\Cscr}{40\Ctenthex}\addtolength
{\Cscr}{-1.0ex}\protect\raisebox{0ex}[40\Ctenthex][\Cscr]{}&\mbox
{if $n\equiv2$ (mod 3),}\\\hspace{3.2ex}0\setlength{\Cscr}{35%
\Ctenthex}\addtolength{\Cscr}{-1.0ex}\protect\raisebox{0ex}[35%
\Ctenthex][\Cscr]{}&\mbox{if $n\equiv0$ (mod 3) and $n\neq0$.}\\%
\end{array}\right.\setlength{\Cscr}{\value{CEht}\Ctenthex}%
\addtolength{\Cscr}{-1.0ex}\protect\raisebox{0ex}[\value{CEht}%
\Ctenthex][\Cscr]{}\protect\end{array}\protect\right.\protect
\label{eq:Exact-fnInfiniteExplicit}\protect\end{eqnarray}%
\setcounter{CEht}{10}We find that \raisebox{0ex}[1ex][0ex]{$%
\protect\displaystyle f_{\:\!\!n\!}$} is an even function of $n$%
; it has a central term of $2/3$, and as we move away from 
\raisebox{0ex}[1ex][0ex]{$\protect\displaystyle n=0$} the terms 
alternate \raisebox{0ex}[1ex][0ex]{$\protect\displaystyle+,-,0,+%
,-,0,\ldots,$} falling off as $1/n$, with a numerical co\-effici%
ent of \raisebox{0ex}[1ex][0ex]{$\protect\displaystyle\sqrt3/2\pi
$}. The sum of all the terms is \setcounter{Ceqindent}{0}\protect
\begin{eqnarray}\hspace{-1.3ex}&\displaystyle\sum_{n=-\infty}^%
\infty\!\!\!f_{\:\!\!n}=\mbox{$\protect\displaystyle\protect\frac
{2}{3}$}+\mbox{$\protect\displaystyle\protect\frac{\sqrt3}{\pi}$%
}\sum_{k=0}^\infty\!\protect\left(\mbox{$\protect\displaystyle
\protect\frac{1}{3k\!+\!1}$}-\mbox{$\protect\displaystyle\protect
\frac{1}{3k\!+\!2}$}\protect\right)\!=1,\protect\nonumber
\setlength{\Cscr}{\value{CEht}\Ctenthex}\addtolength{\Cscr}{-1.0%
ex}\protect\raisebox{0ex}[\value{CEht}\Ctenthex][\Cscr]{}\protect
\end{eqnarray}\setcounter{CEht}{10}as expected: the filter 
\raisebox{0ex}[1ex][0ex]{$\protect\displaystyle f_{\:\!\!n\!}$} 
spreads a delta function out across the lattice without changing 
its integrated value, and so when convolved with any other funct%
ion \raisebox{0ex}[1ex][0ex]{$\protect\displaystyle g_{n\!}$} wi%
ll spread it out somewhat (filtering out the highest momentum st%
ates) without changing its average value.\par Let us now conside%
r obtaining closed-form expressions for \raisebox{0ex}[1ex][0ex]%
{$\protect\displaystyle f_{\:\!\!n}^{N\!}$} for a finite lattice 
of $N$ sites. Since the general method has been described in len%
gthy detail in [\ref{cit:Costella2002a}], I will not repeat it h%
ere, but merely cut directly to the intermediate results.\par Cl%
early, the simplest analysis is for the case when \mbox{$N^{\!}%
\equiv0$ (mod 3)}, because the groups of three successive terms 
in the infinite-lattice result (\protect\ref{eq:Exact-fnInfinite%
Explicit}) will line up in columns. Specifically, all of the 
\mbox{$n\equiv0$ (mod 3)} will fall on every third site, and so 
\raisebox{0ex}[1ex][0ex]{$\protect\displaystyle f_{\:\!\!n\!}$} 
will vanish for all \mbox{$n\equiv0$ (mod 3)} except \raisebox{0%
ex}[1ex][0ex]{$\protect\displaystyle n=0$}, which will retain it%
s infinite-lattice value of $2/3$. For \mbox{$n\not\equiv0$ (mod 
3)}, we obtain the sum \setcounter{Ceqindent}{0}\protect\begin{e%
qnarray}\hspace{-1.3ex}&\displaystyle f_{\mbox{\scriptsize$\:\!%
\!n\!\not\equiv\!0$\,(mod 3)}}^{\mbox{\scriptsize$N\:\!\!\!\equiv
\!0$\,(mod 3)}}=\mbox{$\protect\displaystyle\protect\frac{1}{\pi
}$}\sin\!\protect\left(\mbox{$\protect\displaystyle\protect\frac
{2\pi n}{3}$}\protect\right)\!\sum_{k=0}^\infty\protect\left\{%
\mbox{$\protect\displaystyle\protect\frac{1}{N^{\!}k\!+\!n}$}-%
\mbox{$\protect\displaystyle\protect\frac{1}{N^{\!}k\!+\!N\:\!\!%
\!-\!n}$}\protect\right\}=\mbox{$\protect\displaystyle\protect
\frac{1}{N}$}\sin\!\protect\left(\mbox{$\protect\displaystyle
\protect\frac{2\pi n}{3}$}\protect\right)\cot\!\protect\left(%
\mbox{$\protect\displaystyle\protect\frac{\pi n}{N}$}\protect
\right)\!.\protect\nonumber\setlength{\Cscr}{\value{CEht}%
\Ctenthex}\addtolength{\Cscr}{-1.0ex}\protect\raisebox{0ex}[%
\value{CEht}\Ctenthex][\Cscr]{}\protect\end{eqnarray}\setcounter
{CEht}{10}In actual fact, this expression is also correct for 
\mbox{$n\equiv0$ (mod 3)}, because for \raisebox{0ex}[1ex][0ex]{%
$\protect\displaystyle n\neq0$} the factor \raisebox{0ex}[1ex][0%
ex]{$\protect\displaystyle\sin(2\pi n/3)$} vanishes, and the lim%
it as \raisebox{0ex}[1ex][0ex]{$\protect\displaystyle n%
\rightarrow0$} is the correct value of $2/3$. Thus in full gener%
ality we may write \setcounter{Ceqindent}{0}\protect\begin{eqnar%
ray}\protect\left.\protect\begin{array}{rcl}\protect\displaystyle
\hspace{-1.3ex}&\protect\displaystyle f_{\:\!\!n}^{\mbox{%
\scriptsize$N\:\!\!\!\equiv\!0$\,(mod 3)}}=\mbox{$\protect
\displaystyle\protect\frac{1}{N}$}\sin\!\protect\left(\mbox{$%
\protect\displaystyle\protect\frac{2\pi n}{3}$}\protect\right)%
\cot\!\protect\left(\mbox{$\protect\displaystyle\protect\frac{\pi
n}{N}$}\protect\right)\!,\setlength{\Cscr}{\value{CEht}\Ctenthex
}\addtolength{\Cscr}{-1.0ex}\protect\raisebox{0ex}[\value{CEht}%
\Ctenthex][\Cscr]{}\protect\end{array}\protect\right.\protect
\label{eq:Exact-fnN0mod3}\protect\end{eqnarray}\setcounter{CEht}%
{10}where the limiting procedure for \raisebox{0ex}[1ex][0ex]{$%
\protect\displaystyle n=0$} (yielding the value $2/3$) is to be 
understood.\par For \mbox{$N^{\!}\not\equiv0$ (mod 3)}, the grou%
ps of three terms ``roll through'' every time we loop once aroun%
d the lattice. The simplest way to proceed is to make use of thi%
s $3N$-periodicity rather than the $N$-periodicity of the \mbox{%
$N^{\!}\equiv0$ (mod 3)} case, and to sum up the two different s%
equences.\par Let us start with \mbox{$N^{\!}\equiv1$ (mod 3)}. 
The \raisebox{0ex}[1ex][0ex]{$\protect\displaystyle n=0$} term p%
icks up a correction: \setcounter{Ceqindent}{0}\protect\begin{eq%
narray}\protect\left.\protect\begin{array}{rcl}\protect
\displaystyle\hspace{-1.3ex}&\protect\displaystyle f_{\:\!\!n=0}%
^{\mbox{\scriptsize$N\:\!\!\!\equiv\!1$\,(mod 3)}}=\mbox{$%
\protect\displaystyle\protect\frac{2}{3}$}+\mbox{$\protect
\displaystyle\protect\frac{\sqrt3}{\pi}$}\sum_{k=0}^\infty
\protect\left\{\mbox{$\protect\displaystyle\protect\frac{1}{3N^{%
\!}k\!+\!N}$}-\mbox{$\protect\displaystyle\protect\frac{1}{3N^{%
\!}k\!+\!2N}$}\protect\right\}=\mbox{$\protect\displaystyle
\protect\frac{2}{3}$}\!\protect\left(1+\mbox{$\protect
\displaystyle\protect\frac{1}{2N}$}^{\!}\protect\right)\!.%
\setlength{\Cscr}{\value{CEht}\Ctenthex}\addtolength{\Cscr}{-1.0%
ex}\protect\raisebox{0ex}[\value{CEht}\Ctenthex][\Cscr]{}\protect
\end{array}\protect\right.\protect\label{eq:Exact-fnN1mod3n0}%
\protect\end{eqnarray}\setcounter{CEht}{10}For \mbox{$n\equiv1$ 
(mod 3)} we obtain \setcounter{Ceqindent}{0}\protect\begin{eqnar%
ray}f_{\mbox{\scriptsize$\:\!\!n\!\equiv\!1$\,(mod 3)}}^{\mbox{%
\scriptsize$N\:\!\!\!\equiv\!1$\,(mod 3)}}\hspace{-1.3ex}&%
\displaystyle=&\hspace{-1.3ex}\mbox{$\protect\displaystyle
\protect\frac{\sqrt3}{2\pi}$}\sum_{k=0}^\infty\setcounter{Cbscur%
r}{30}\setlength{\Cscr}{\value{Cbscurr}\Ctenthex}\addtolength{%
\Cscr}{-1.0ex}\protect\raisebox{0ex}[\value{Cbscurr}\Ctenthex][%
\Cscr]{}\hspace{-0ex}{\protect\left\{\setlength{\Cscr}{\value{Cb%
scurr}\Ctenthex}\addtolength{\Cscr}{-1.0ex}\protect\raisebox{0ex%
}[\value{Cbscurr}\Ctenthex][\Cscr]{}\protect\right.}\hspace{-0.2%
5ex}\setlength{\Cscr}{\value{Cbscurr}\Ctenthex}\addtolength{\Cscr
}{-1.0ex}\protect\raisebox{0ex}[\value{Cbscurr}\Ctenthex][\Cscr]%
{}\setcounter{CbsD}{\value{CbsC}}\setcounter{CbsC}{\value{CbsB}}%
\setcounter{CbsB}{\value{CbsA}}\setcounter{CbsA}{\value{Cbscurr}%
}\!\!\protect\left(\mbox{$\protect\displaystyle\protect\frac{1}{%
3N^{\!}k\!+\!n}$}-\mbox{$\protect\displaystyle\protect\frac{1}{3%
N^{\!}k\!+\!3N\:\!\!\!-\!n}$}\protect\right)\setcounter{Ceqinden%
t}{250}\protect\nonumber\setlength{\Cscr}{\value{CEht}\Ctenthex}%
\addtolength{\Cscr}{-1.0ex}\protect\raisebox{0ex}[\value{CEht}%
\Ctenthex][\Cscr]{}\\*[0ex]\protect\displaystyle\hspace{-1.3ex}&%
\displaystyle&\hspace{-1.3ex}{\protect\mbox{}}\hspace{\value{Ceq%
indent}\Ctenthex}-\protect\left(\mbox{$\protect\displaystyle
\protect\frac{1}{3N^{\!}k\!+\!(N\!+\!n)}$}-\mbox{$\protect
\displaystyle\protect\frac{1}{3N^{\!}k\!+\!3N\:\!\!\!-\!(N\!+\!n%
)}$}\protect\right)\!\!\setlength{\Cscr}{\value{CbsA}\Ctenthex}%
\addtolength{\Cscr}{-1.0ex}\protect\raisebox{0ex}[\value{CbsA}%
\Ctenthex][\Cscr]{}\hspace{-0.25ex}{\protect\left.\setlength{%
\Cscr}{\value{CbsA}\Ctenthex}\addtolength{\Cscr}{-1.0ex}\protect
\raisebox{0ex}[\value{CbsA}\Ctenthex][\Cscr]{}\protect\right\}}%
\hspace{-0ex}\setlength{\Cscr}{\value{CbsA}\Ctenthex}\addtolength
{\Cscr}{-1.0ex}\protect\raisebox{0ex}[\value{CbsA}\Ctenthex][%
\Cscr]{}\setcounter{CbsA}{\value{CbsB}}\setcounter{CbsB}{\value{%
CbsC}}\setcounter{CbsC}{\value{CbsD}}\setcounter{CbsD}{1}\protect
\nonumber\setlength{\Cscr}{\value{CEht}\Ctenthex}\addtolength{%
\Cscr}{-1.0ex}\protect\raisebox{0ex}[\value{CEht}\Ctenthex][\Cscr
]{}\\*[0ex]\protect\displaystyle\hspace{-1.3ex}&\displaystyle=&%
\hspace{-1.3ex}\mbox{$\protect\displaystyle\protect\frac{1}{2N^{%
\!}\sqrt3}$}\:\!\!\protect\left\{\cot\!\protect\left(\mbox{$%
\protect\displaystyle\protect\frac{\pi n}{3N}$}\protect\right)^{%
\!}-\cot\!\protect\left(\mbox{$\protect\displaystyle\protect\frac
{\pi}{3}$}\!+\!\mbox{$\protect\displaystyle\protect\frac{\pi n}{%
3N}$}\protect\right)\!\protect\right\}\!.\protect\nonumber
\setlength{\Cscr}{\value{CEht}\Ctenthex}\addtolength{\Cscr}{-1.0%
ex}\protect\raisebox{0ex}[\value{CEht}\Ctenthex][\Cscr]{}\protect
\end{eqnarray}\setcounter{CEht}{10}For \mbox{$n\equiv2$ (mod 3)} 
we obtain \setcounter{Ceqindent}{0}\protect\begin{eqnarray}f_{%
\mbox{\scriptsize$\:\!\!n\!\equiv\!2$\,(mod 3)}}^{\mbox{%
\scriptsize$N\:\!\!\!\equiv\!1$\,(mod 3)}}\hspace{-1.3ex}&%
\displaystyle=&\hspace{-1.3ex}\mbox{$\protect\displaystyle
\protect\frac{\sqrt3}{2\pi}$}\sum_{k=0}^\infty\setcounter{Cbscur%
r}{30}\setlength{\Cscr}{\value{Cbscurr}\Ctenthex}\addtolength{%
\Cscr}{-1.0ex}\protect\raisebox{0ex}[\value{Cbscurr}\Ctenthex][%
\Cscr]{}\hspace{-0ex}{\protect\left\{\setlength{\Cscr}{\value{Cb%
scurr}\Ctenthex}\addtolength{\Cscr}{-1.0ex}\protect\raisebox{0ex%
}[\value{Cbscurr}\Ctenthex][\Cscr]{}\protect\right.}\hspace{-0.2%
5ex}\setlength{\Cscr}{\value{Cbscurr}\Ctenthex}\addtolength{\Cscr
}{-1.0ex}\protect\raisebox{0ex}[\value{Cbscurr}\Ctenthex][\Cscr]%
{}\setcounter{CbsD}{\value{CbsC}}\setcounter{CbsC}{\value{CbsB}}%
\setcounter{CbsB}{\value{CbsA}}\setcounter{CbsA}{\value{Cbscurr}%
}\!\!\protect\left(\mbox{$\protect\displaystyle\protect\frac{1}{%
3N^{\!}k\!+\!(2N\:\!\!\!+\!n)}$}-\mbox{$\protect\displaystyle
\protect\frac{1}{3N^{\!}k\!+\!3N\:\!\!\!-\!(2N\:\!\!\!+\!n)}$}%
\protect\right)\setcounter{Ceqindent}{290}\protect\nonumber
\setlength{\Cscr}{\value{CEht}\Ctenthex}\addtolength{\Cscr}{-1.0%
ex}\protect\raisebox{0ex}[\value{CEht}\Ctenthex][\Cscr]{}\\*[0ex%
]\protect\displaystyle\hspace{-1.3ex}&\displaystyle&\hspace{-1.3%
ex}{\protect\mbox{}}\hspace{\value{Ceqindent}\Ctenthex}-\protect
\left(\mbox{$\protect\displaystyle\protect\frac{1}{3N^{\!}k\!+\!%
n}$}-\mbox{$\protect\displaystyle\protect\frac{1}{3N^{\!}k\!+\!3%
N\:\!\!\!-\!n}$}\protect\right)\!\!\setlength{\Cscr}{\value{CbsA%
}\Ctenthex}\addtolength{\Cscr}{-1.0ex}\protect\raisebox{0ex}[%
\value{CbsA}\Ctenthex][\Cscr]{}\hspace{-0.25ex}{\protect\left.%
\setlength{\Cscr}{\value{CbsA}\Ctenthex}\addtolength{\Cscr}{-1.0%
ex}\protect\raisebox{0ex}[\value{CbsA}\Ctenthex][\Cscr]{}\protect
\right\}}\hspace{-0ex}\setlength{\Cscr}{\value{CbsA}\Ctenthex}%
\addtolength{\Cscr}{-1.0ex}\protect\raisebox{0ex}[\value{CbsA}%
\Ctenthex][\Cscr]{}\setcounter{CbsA}{\value{CbsB}}\setcounter{Cb%
sB}{\value{CbsC}}\setcounter{CbsC}{\value{CbsD}}\setcounter{CbsD%
}{1}\protect\nonumber\setlength{\Cscr}{\value{CEht}\Ctenthex}%
\addtolength{\Cscr}{-1.0ex}\protect\raisebox{0ex}[\value{CEht}%
\Ctenthex][\Cscr]{}\\*[0ex]\protect\displaystyle\hspace{-1.3ex}&%
\displaystyle=&\hspace{-1.3ex}\mbox{$\protect\displaystyle
\protect\frac{1}{2N^{\!}\sqrt3}$}\:\!\!\protect\left\{\cot\!%
\protect\left(\mbox{$\protect\displaystyle\protect\frac{2\pi}{3}%
$}\!+\!\mbox{$\protect\displaystyle\protect\frac{\pi n}{3N}$}%
\protect\right)^{\!}-\cot\!\protect\left(\mbox{$\protect
\displaystyle\protect\frac{\pi n}{3N}$}\protect\right)\!\protect
\right\}\!.\protect\nonumber\setlength{\Cscr}{\value{CEht}%
\Ctenthex}\addtolength{\Cscr}{-1.0ex}\protect\raisebox{0ex}[%
\value{CEht}\Ctenthex][\Cscr]{}\protect\end{eqnarray}\setcounter
{CEht}{10}Finally, for \mbox{$n\equiv0$ (mod 3)}, \raisebox{0ex}%
[1ex][0ex]{$\protect\displaystyle n\neq0$}, we obtain \setcounter
{Ceqindent}{0}\protect\begin{eqnarray}f_{\mbox{\scriptsize$\:\!%
\!n\!\equiv\!0$\,(mod 3)}}^{\mbox{\scriptsize$N\:\!\!\!\equiv\!1%
$\,(mod 3)}}\hspace{-1.3ex}&\displaystyle=&\hspace{-1.3ex}\mbox{%
$\protect\displaystyle\protect\frac{\sqrt3}{2\pi}$}\sum_{k=0}^%
\infty\setcounter{Cbscurr}{30}\setlength{\Cscr}{\value{Cbscurr}%
\Ctenthex}\addtolength{\Cscr}{-1.0ex}\protect\raisebox{0ex}[%
\value{Cbscurr}\Ctenthex][\Cscr]{}\hspace{-0ex}{\protect\left\{%
\setlength{\Cscr}{\value{Cbscurr}\Ctenthex}\addtolength{\Cscr}{-%
1.0ex}\protect\raisebox{0ex}[\value{Cbscurr}\Ctenthex][\Cscr]{}%
\protect\right.}\hspace{-0.25ex}\setlength{\Cscr}{\value{Cbscurr%
}\Ctenthex}\addtolength{\Cscr}{-1.0ex}\protect\raisebox{0ex}[%
\value{Cbscurr}\Ctenthex][\Cscr]{}\setcounter{CbsD}{\value{CbsC}%
}\setcounter{CbsC}{\value{CbsB}}\setcounter{CbsB}{\value{CbsA}}%
\setcounter{CbsA}{\value{Cbscurr}}\!\!\protect\left(\mbox{$%
\protect\displaystyle\protect\frac{1}{3N^{\!}k\!+\!(N\:\!\!\!+n)%
}$}-\mbox{$\protect\displaystyle\protect\frac{1}{3N^{\!}k\!+\!3N%
\:\!\!\!-\!(N\:\!\!\!+n)}$}\protect\right)\setcounter{Ceqindent}%
{250}\protect\nonumber\setlength{\Cscr}{\value{CEht}\Ctenthex}%
\addtolength{\Cscr}{-1.0ex}\protect\raisebox{0ex}[\value{CEht}%
\Ctenthex][\Cscr]{}\\*[0ex]\protect\displaystyle\hspace{-1.3ex}&%
\displaystyle&\hspace{-1.3ex}{\protect\mbox{}}\hspace{\value{Ceq%
indent}\Ctenthex}-\protect\left(\mbox{$\protect\displaystyle
\protect\frac{1}{3N^{\!}k\!+\!(2N\:\!\!\!+\!n)}$}-\mbox{$\protect
\displaystyle\protect\frac{1}{3N^{\!}k\!+\!3N\:\!\!\!-\!(2N\:\!%
\!\!+\!n)}$}\protect\right)\!\!\setlength{\Cscr}{\value{CbsA}%
\Ctenthex}\addtolength{\Cscr}{-1.0ex}\protect\raisebox{0ex}[%
\value{CbsA}\Ctenthex][\Cscr]{}\hspace{-0.25ex}{\protect\left.%
\setlength{\Cscr}{\value{CbsA}\Ctenthex}\addtolength{\Cscr}{-1.0%
ex}\protect\raisebox{0ex}[\value{CbsA}\Ctenthex][\Cscr]{}\protect
\right\}}\hspace{-0ex}\setlength{\Cscr}{\value{CbsA}\Ctenthex}%
\addtolength{\Cscr}{-1.0ex}\protect\raisebox{0ex}[\value{CbsA}%
\Ctenthex][\Cscr]{}\setcounter{CbsA}{\value{CbsB}}\setcounter{Cb%
sB}{\value{CbsC}}\setcounter{CbsC}{\value{CbsD}}\setcounter{CbsD%
}{1}\protect\nonumber\setlength{\Cscr}{\value{CEht}\Ctenthex}%
\addtolength{\Cscr}{-1.0ex}\protect\raisebox{0ex}[\value{CEht}%
\Ctenthex][\Cscr]{}\\*[0ex]\protect\displaystyle\hspace{-1.3ex}&%
\displaystyle=&\hspace{-1.3ex}\mbox{$\protect\displaystyle
\protect\frac{1}{2N^{\!}\sqrt3}$}\:\!\!\protect\left\{\cot\!%
\protect\left(\mbox{$\protect\displaystyle\protect\frac{\pi}{3}$%
}\!+\!\mbox{$\protect\displaystyle\protect\frac{\pi n}{3N}$}%
\protect\right)^{\!}-\cot\!\protect\left(\mbox{$\protect
\displaystyle\protect\frac{2\pi}{3}$}\!+\!\mbox{$\protect
\displaystyle\protect\frac{\pi n}{3N}$}\protect\right)\!\protect
\right\}\!,\protect\nonumber\setlength{\Cscr}{\value{CEht}%
\Ctenthex}\addtolength{\Cscr}{-1.0ex}\protect\raisebox{0ex}[%
\value{CEht}\Ctenthex][\Cscr]{}\protect\end{eqnarray}\setcounter
{CEht}{10}We can now see that there are three sequences interlea%
ved here: one with the cot of \raisebox{0ex}[1ex][0ex]{$\protect
\displaystyle\pi n/3N$}, one with $\pi/3$ added to the angle, an%
d one with $2\pi/3$ added to the angle. We can thus combine them 
back into a single expression: \setcounter{Ceqindent}{0}\protect
\begin{eqnarray}\protect\left.\protect\begin{array}{rcl}\protect
\displaystyle f_{\:\!\!n}^{\mbox{\scriptsize$N\:\!\!\!\equiv\!1$%
\,(mod 3)}}\hspace{-1.3ex}&\protect\displaystyle=&\hspace{-1.3ex%
}\protect\displaystyle\mbox{$\protect\displaystyle\protect\frac{%
1}{3N}$}\setcounter{Cbscurr}{30}\setlength{\Cscr}{\value{Cbscurr%
}\Ctenthex}\addtolength{\Cscr}{-1.0ex}\protect\raisebox{0ex}[%
\value{Cbscurr}\Ctenthex][\Cscr]{}\hspace{-0ex}{\protect\left\{%
\setlength{\Cscr}{\value{Cbscurr}\Ctenthex}\addtolength{\Cscr}{-%
1.0ex}\protect\raisebox{0ex}[\value{Cbscurr}\Ctenthex][\Cscr]{}%
\protect\right.}\hspace{-0.25ex}\setlength{\Cscr}{\value{Cbscurr%
}\Ctenthex}\addtolength{\Cscr}{-1.0ex}\protect\raisebox{0ex}[%
\value{Cbscurr}\Ctenthex][\Cscr]{}\setcounter{CbsD}{\value{CbsC}%
}\setcounter{CbsC}{\value{CbsB}}\setcounter{CbsB}{\value{CbsA}}%
\setcounter{CbsA}{\value{Cbscurr}}\sin\!\protect\left(\mbox{$%
\protect\displaystyle\protect\frac{2\pi n}{3}$}\protect\right)%
\cot\!\protect\left(\mbox{$\protect\displaystyle\protect\frac{\pi
n}{3N}$}\protect\right)^{\!}+\sin\!\protect\left(\mbox{$\protect
\displaystyle\protect\frac{2\pi}{3}$}\!+\!\mbox{$\protect
\displaystyle\protect\frac{2\pi n}{3}$}\protect\right)^{\!}\cot
\!\protect\left(\mbox{$\protect\displaystyle\protect\frac{\pi}{3%
}$}\!+\!\mbox{$\protect\displaystyle\protect\frac{\pi n}{3N}$}%
\protect\right)^{\!}\setcounter{Ceqindent}{350}\setlength{\Cscr}%
{\value{CEht}\Ctenthex}\addtolength{\Cscr}{-1.0ex}\protect
\raisebox{0ex}[\value{CEht}\Ctenthex][\Cscr]{}\\*[0.55ex]\protect
\displaystyle\hspace{-1.3ex}&\protect\displaystyle&\hspace{-1.3e%
x}\protect\displaystyle{\protect\mbox{}}\hspace{\value{Ceqindent%
}\Ctenthex}+\sin\!\protect\left(\mbox{$\protect\displaystyle
\protect\frac{4\pi}{3}$}\!+\!\mbox{$\protect\displaystyle\protect
\frac{2\pi n}{3}$}\protect\right)^{\!}\cot\!\protect\left(\mbox{%
$\protect\displaystyle\protect\frac{2\pi}{3}$}\!+\!\mbox{$%
\protect\displaystyle\protect\frac{\pi n}{3N}$}\protect\right)\!%
\!\setlength{\Cscr}{\value{CbsA}\Ctenthex}\addtolength{\Cscr}{-1%
.0ex}\protect\raisebox{0ex}[\value{CbsA}\Ctenthex][\Cscr]{}%
\hspace{-0.25ex}{\protect\left.\setlength{\Cscr}{\value{CbsA}%
\Ctenthex}\addtolength{\Cscr}{-1.0ex}\protect\raisebox{0ex}[%
\value{CbsA}\Ctenthex][\Cscr]{}\protect\right\}}\hspace{-0ex}%
\setlength{\Cscr}{\value{CbsA}\Ctenthex}\addtolength{\Cscr}{-1.0%
ex}\protect\raisebox{0ex}[\value{CbsA}\Ctenthex][\Cscr]{}%
\setcounter{CbsA}{\value{CbsB}}\setcounter{CbsB}{\value{CbsC}}%
\setcounter{CbsC}{\value{CbsD}}\setcounter{CbsD}{1},\setlength{%
\Cscr}{\value{CEht}\Ctenthex}\addtolength{\Cscr}{-1.0ex}\protect
\raisebox{0ex}[\value{CEht}\Ctenthex][\Cscr]{}\protect\end{array%
}\protect\right.\protect\label{eq:Exact-fnN1mod3}\protect\end{eq%
narray}\setcounter{CEht}{10}where again we find that the 
\raisebox{0ex}[1ex][0ex]{$\protect\displaystyle n=0$} term is au%
tomatically included as the \raisebox{0ex}[1ex][0ex]{$\protect
\displaystyle n\rightarrow0$} limit of the generic expression.%
\par Finally, we can carry through the same analysis for \mbox{$%
N^{\!}\equiv2$ (mod 3)}. When all the terms are recombined, the 
only difference from (\protect\ref{eq:Exact-fnN1mod3}) is that t%
he ``rolling'' factor ``rolls'' in the reverse direction: 
\setcounter{Ceqindent}{0}\protect\begin{eqnarray}\protect\left.%
\protect\begin{array}{rcl}\protect\displaystyle f_{\:\!\!n}^{%
\mbox{\scriptsize$N\:\!\!\!\equiv\!2$\,(mod 3)}}\hspace{-1.3ex}&%
\protect\displaystyle=&\hspace{-1.3ex}\protect\displaystyle\mbox
{$\protect\displaystyle\protect\frac{1}{3N}$}\setcounter{Cbscurr%
}{30}\setlength{\Cscr}{\value{Cbscurr}\Ctenthex}\addtolength{%
\Cscr}{-1.0ex}\protect\raisebox{0ex}[\value{Cbscurr}\Ctenthex][%
\Cscr]{}\hspace{-0ex}{\protect\left\{\setlength{\Cscr}{\value{Cb%
scurr}\Ctenthex}\addtolength{\Cscr}{-1.0ex}\protect\raisebox{0ex%
}[\value{Cbscurr}\Ctenthex][\Cscr]{}\protect\right.}\hspace{-0.2%
5ex}\setlength{\Cscr}{\value{Cbscurr}\Ctenthex}\addtolength{\Cscr
}{-1.0ex}\protect\raisebox{0ex}[\value{Cbscurr}\Ctenthex][\Cscr]%
{}\setcounter{CbsD}{\value{CbsC}}\setcounter{CbsC}{\value{CbsB}}%
\setcounter{CbsB}{\value{CbsA}}\setcounter{CbsA}{\value{Cbscurr}%
}\sin\!\protect\left(\mbox{$\protect\displaystyle\protect\frac{2%
\pi n}{3}$}\protect\right)\cot\!\protect\left(\mbox{$\protect
\displaystyle\protect\frac{\pi n}{3N}$}\protect\right)^{\!}+\sin
\!\protect\left(\mbox{$\protect\displaystyle\protect\frac{4\pi}{%
3}$}\!+\!\mbox{$\protect\displaystyle\protect\frac{2\pi n}{3}$}%
\protect\right)^{\!}\cot\!\protect\left(\mbox{$\protect
\displaystyle\protect\frac{\pi}{3}$}\!+\!\mbox{$\protect
\displaystyle\protect\frac{\pi n}{3N}$}\protect\right)^{\!}%
\setcounter{Ceqindent}{350}\setlength{\Cscr}{\value{CEht}%
\Ctenthex}\addtolength{\Cscr}{-1.0ex}\protect\raisebox{0ex}[%
\value{CEht}\Ctenthex][\Cscr]{}\\*[0.55ex]\protect\displaystyle
\hspace{-1.3ex}&\protect\displaystyle&\hspace{-1.3ex}\protect
\displaystyle{\protect\mbox{}}\hspace{\value{Ceqindent}\Ctenthex
}+\sin\!\protect\left(\mbox{$\protect\displaystyle\protect\frac{%
2\pi}{3}$}\!+\!\mbox{$\protect\displaystyle\protect\frac{2\pi n}%
{3}$}\protect\right)^{\!}\cot\!\protect\left(\mbox{$\protect
\displaystyle\protect\frac{2\pi}{3}$}\!+\!\mbox{$\protect
\displaystyle\protect\frac{\pi n}{3N}$}\protect\right)\!\!%
\setlength{\Cscr}{\value{CbsA}\Ctenthex}\addtolength{\Cscr}{-1.0%
ex}\protect\raisebox{0ex}[\value{CbsA}\Ctenthex][\Cscr]{}\hspace
{-0.25ex}{\protect\left.\setlength{\Cscr}{\value{CbsA}\Ctenthex}%
\addtolength{\Cscr}{-1.0ex}\protect\raisebox{0ex}[\value{CbsA}%
\Ctenthex][\Cscr]{}\protect\right\}}\hspace{-0ex}\setlength{\Cscr
}{\value{CbsA}\Ctenthex}\addtolength{\Cscr}{-1.0ex}\protect
\raisebox{0ex}[\value{CbsA}\Ctenthex][\Cscr]{}\setcounter{CbsA}{%
\value{CbsB}}\setcounter{CbsB}{\value{CbsC}}\setcounter{CbsC}{%
\value{CbsD}}\setcounter{CbsD}{1}.\setlength{\Cscr}{\value{CEht}%
\Ctenthex}\addtolength{\Cscr}{-1.0ex}\protect\raisebox{0ex}[%
\value{CEht}\Ctenthex][\Cscr]{}\protect\end{array}\protect\right
.\protect\label{eq:Exact-fnN2mod3}\protect\end{eqnarray}%
\setcounter{CEht}{10}Again, we find that the \raisebox{0ex}[1ex]%
[0ex]{$\protect\displaystyle n=0$} term is included as the 
\raisebox{0ex}[1ex][0ex]{$\protect\displaystyle n\rightarrow0$} 
limit of the generic expression, in agreement with the first-pri%
nciples result: \setcounter{Ceqindent}{0}\protect\begin{eqnarray%
}\protect\left.\protect\begin{array}{rcl}\protect\displaystyle
\hspace{-1.3ex}&\protect\displaystyle f_{\:\!\!n=0}^{\mbox{%
\scriptsize$N\:\!\!\!\equiv\!2$\,(mod 3)}}=\mbox{$\protect
\displaystyle\protect\frac{2}{3}$}+\mbox{$\protect\displaystyle
\protect\frac{\sqrt3}{\pi}$}\sum_{k=0}^\infty\protect\left\{\mbox
{$\protect\displaystyle\protect\frac{1}{3N^{\!}k\!+\!2N}$}-\mbox
{$\protect\displaystyle\protect\frac{1}{3N^{\!}k\!+\!N}$}\protect
\right\}=\mbox{$\protect\displaystyle\protect\frac{2}{3}$}\!%
\protect\left(1-\mbox{$\protect\displaystyle\protect\frac{1}{2N}%
$}^{\!}\protect\right)\!.\setlength{\Cscr}{\value{CEht}\Ctenthex
}\addtolength{\Cscr}{-1.0ex}\protect\raisebox{0ex}[\value{CEht}%
\Ctenthex][\Cscr]{}\protect\end{array}\protect\right.\protect
\label{eq:Exact-fnN2mod3n0}\protect\end{eqnarray}\setcounter{CEh%
t}{10}\par Finally, we can combine together the results (\protect
\ref{eq:Exact-fnN1mod3}) and (\protect\ref{eq:Exact-fnN2mod3}) f%
or \mbox{$N^{\!}\not\equiv0$ (mod 3)}, by means of a simple fact%
or in the argument of the ``rolling factor'' that depends on the 
value of \mbox{$k\equiv N\:\!\!$ (mod 3)}: \setcounter{Ceqindent%
}{0}\protect\begin{eqnarray}\protect\left.\protect\begin{array}{%
rcl}\protect\displaystyle f_{\:\!\!n}^{\mbox{\scriptsize$N\:\!\!%
\!\equiv\!k\!\neq\!0$\,(mod 3)}}\hspace{-1.3ex}&\protect
\displaystyle=&\hspace{-1.3ex}\protect\displaystyle\mbox{$%
\protect\displaystyle\protect\frac{1}{3N}$}\setcounter{Cbscurr}{%
30}\setlength{\Cscr}{\value{Cbscurr}\Ctenthex}\addtolength{\Cscr
}{-1.0ex}\protect\raisebox{0ex}[\value{Cbscurr}\Ctenthex][\Cscr]%
{}\hspace{-0ex}{\protect\left\{\setlength{\Cscr}{\value{Cbscurr}%
\Ctenthex}\addtolength{\Cscr}{-1.0ex}\protect\raisebox{0ex}[%
\value{Cbscurr}\Ctenthex][\Cscr]{}\protect\right.}\hspace{-0.25e%
x}\setlength{\Cscr}{\value{Cbscurr}\Ctenthex}\addtolength{\Cscr}%
{-1.0ex}\protect\raisebox{0ex}[\value{Cbscurr}\Ctenthex][\Cscr]{%
}\setcounter{CbsD}{\value{CbsC}}\setcounter{CbsC}{\value{CbsB}}%
\setcounter{CbsB}{\value{CbsA}}\setcounter{CbsA}{\value{Cbscurr}%
}\sin\!\protect\left(\mbox{$\protect\displaystyle\protect\frac{2%
\pi n}{3}$}\protect\right)\cot\!\protect\left(\mbox{$\protect
\displaystyle\protect\frac{\pi n}{3N}$}\protect\right)^{\!}+\sin
\!\protect\left(\mbox{$\protect\displaystyle\protect\frac{2k\pi}%
{3}$}\!+\!\mbox{$\protect\displaystyle\protect\frac{2\pi n}{3}$}%
\protect\right)^{\!}\cot\!\protect\left(\mbox{$\protect
\displaystyle\protect\frac{\pi}{3}$}\!+\!\mbox{$\protect
\displaystyle\protect\frac{\pi n}{3N}$}\protect\right)^{\!}%
\setcounter{Ceqindent}{310}\setlength{\Cscr}{\value{CEht}%
\Ctenthex}\addtolength{\Cscr}{-1.0ex}\protect\raisebox{0ex}[%
\value{CEht}\Ctenthex][\Cscr]{}\\*[0.55ex]\protect\displaystyle
\hspace{-1.3ex}&\protect\displaystyle&\hspace{-1.3ex}\protect
\displaystyle{\protect\mbox{}}\hspace{\value{Ceqindent}\Ctenthex
}+\sin\!\protect\left(\mbox{$\protect\displaystyle\protect\frac{%
4k\pi}{3}$}\!+\!\mbox{$\protect\displaystyle\protect\frac{2\pi n%
}{3}$}\protect\right)^{\!}\cot\!\protect\left(\mbox{$\protect
\displaystyle\protect\frac{2\pi}{3}$}\!+\!\mbox{$\protect
\displaystyle\protect\frac{\pi n}{3N}$}\protect\right)\!\!%
\setlength{\Cscr}{\value{CbsA}\Ctenthex}\addtolength{\Cscr}{-1.0%
ex}\protect\raisebox{0ex}[\value{CbsA}\Ctenthex][\Cscr]{}\hspace
{-0.25ex}{\protect\left.\setlength{\Cscr}{\value{CbsA}\Ctenthex}%
\addtolength{\Cscr}{-1.0ex}\protect\raisebox{0ex}[\value{CbsA}%
\Ctenthex][\Cscr]{}\protect\right\}}\hspace{-0ex}\setlength{\Cscr
}{\value{CbsA}\Ctenthex}\addtolength{\Cscr}{-1.0ex}\protect
\raisebox{0ex}[\value{CbsA}\Ctenthex][\Cscr]{}\setcounter{CbsA}{%
\value{CbsB}}\setcounter{CbsB}{\value{CbsC}}\setcounter{CbsC}{%
\value{CbsD}}\setcounter{CbsD}{1}.\setlength{\Cscr}{\value{CEht}%
\Ctenthex}\addtolength{\Cscr}{-1.0ex}\protect\raisebox{0ex}[%
\value{CEht}\Ctenthex][\Cscr]{}\protect\end{array}\protect\right
.\protect\label{eq:Exact-fnNnot0mod3}\protect\end{eqnarray}%
\setcounter{CEht}{10}\par The expressions (\protect\ref{eq:Exact%
-fnN0mod3}) and (\protect\ref{eq:Exact-fnNnot0mod3}) are the fin%
al results of our calculation. These weights may be calculated f%
or any value of the lattice size $N$, stored, and used to provid%
e an antialiasing filter in that dimension. For a multi-dimensio%
nal lattice, we need simply perform the filtering operation in e%
ach dimension separately. If the dimensions have different sizes%
, then we need to calculate and store a separate set of weights 
for each different size. Apart from that, however, the weights r%
emain fixed for the entire run of the lattice calculation.\par
\refstepcounter{section}\vspace{1.5\baselineskip}\par{\centering
\bf\thesection. Stochastic implementation of the antialiasing fi%
lter\\*[0.5\baselineskip]}\protect\indent\label{sect:Stochastic}%
Clearly, the antialiasing filter described in the previous secti%
on is not suitable for direct implementation in practical lattic%
e calculations. It is as horrifically nonlocal as the SLAC deriv%
ative operator studied in [\ref{cit:Costella2002a},\,\ref{cit:Co%
stella2002b},\,\ref{cit:Costella2004}], linking a given site to 
every other site along the dimension of the lattice that we are 
filtering, with the absolute value of each term falling off only 
inversely with distance in the infinite-lattice case (with small 
numerical corrections for finite lattices). This means that it i%
s not possible to truncate the series at some finite distance wi%
thout severely distorting the fidelity of the operator (like we 
can with operators that fall off exponentially with distance)---%
which is precisely the fate that we want to avoid.\par Thus, any 
practical implementation of the antialiasing filter would almost 
certainly employ the stochastic approach suggested in [\ref{cit:%
Costella2002a}] for the SLAC operators. Namely, the absolute val%
ue of each weight calculated in the previous section will be tak%
en to be the \mbox{}\protect\/{\protect\em probability\protect\/%
} that the given linkage will be made between the site at which 
the antialiasing filter is being applied and the site which is a%
t the given distance from the said site. Since the magnitude of 
the terms fall off inversely with distance, this implies that th%
e average number of linkages that will be performed per site wil%
l be proportional to the \mbox{}\protect\/{\protect\em logarithm%
\protect\/} of the length of the lattice in the given dimension, 
rather than being proportional to the length itself.\par This is 
precisely the philosophy espoused in [\ref{cit:Costella2002a},\,%
\ref{cit:Costella2002b},\,\ref{cit:Costella2004}], and so does n%
ot need to be elaborated on further here. However, I need to poi%
nt out one important practical error contained in those previous 
papers, and one improvement in the algorithm that is of particul%
ar interest to the current case.\par The error relates to the re%
quirement that each and every step of a lattice calculation main%
tain the unitarity of the formalism. In each of [\ref{cit:Costel%
la2002a},\,\ref{cit:Costella2002b},\,\ref{cit:Costella2004}]\ I 
either stated or implied that unitarity would be maintained if e%
very computation of a positive-distance term from a given site w%
as always accompanied by the computation of the corresponding ne%
gative-distance term from that given site. This erroneous claim 
was based on a misapplication of the requirements of Hermiticity 
of the momentum operator. When the Hermiticity requirements are 
examined in closer detail, what we find is something that looks 
somewhat similar at a glance, but which has completely different 
ramifications for practical calculations, namely, that if for a 
Hermitian operator we perform a calculation at any site $i$ that 
depends on the value of a field at any other site $j$, then we 
\mbox{}\protect\/{\protect\em must\protect\/} perform the Hermit%
ian-conjugate calculation at site $j$ that depends on the Hermit%
ian-conjugate value of that same field at site $i$. In other wor%
ds, we have a form of ``reciprocity requirement'': if a site ``g%
ives''\mbox{$\!$}, then it must ``receive''\mbox{$\!$}. It is 
\mbox{}\protect\/{\protect\em not\protect\/} at all required tha%
t a positive-distance calculation from a given site be always ac%
companied by a negative-distance calculation from that same site 
(although this will be true on the average).\par The algorithmic 
improvement relates to the implementation of the probabilistic a%
spect of each possible ``linkage'' between sites. In [\ref{cit:C%
ostella2002a}] I pointed out that the evaluation of a probabilit%
y of $1/n$ (with $n$ an integer) can be carried out much more ef%
ficiently than for an arbitrary real probability between 0 and 1%
, because the former can be determined directly from the pseudo-%
random bitstream, without the need for floating-point calculatio%
ns at all. This was very fortuitous for the considerations of [%
\ref{cit:Costella2002a}], because the absolute value of each ter%
m of the first-derivative operator for an infinite lattice is pr%
ecisely $1/n$, where $n$ is the distance in lattice sites. This 
aspect of my argument remains true.\par However, when in [\ref{c%
it:Costella2002b}] I ``corrected'' the infinite-lattice results 
for finite lattices, the absolute value for the co\-efficient fo%
r a distance of $n$ was changed from $1/n$ to some other number 
that, for small $n$, was very close to $1/n$. I therefore propos%
ed that the probability for performing each calculation be maint%
ained at $1/n$, because this yes/no decision needs to be perform%
ed as quickly as possible, as it performed for \mbox{}\protect\/%
{\protect\em every\protect\/} distance from the given site. Rath%
er, \mbox{}\protect\/{\protect\em if\protect\/} the result of th%
is probabilistic decision was to actually perform the linkage, t%
hen the calculation would be multiplied by a factor that was the 
ratio of the true (corrected) weight to $1/n$. The argument goes 
that, since these ``correction factors'' are all close to unity, 
the optimal balance obtained in [\ref{cit:Costella2002a}] betwee%
n the statistical noise of the operator and the minimisation of 
the number of calculations is effectively maintained, without th%
e need to introduce floating-point calculations into the innermo%
st probabilistic decision loop.\par The logic of this argument a%
lso remains true. However, what I overlooked at the time was the 
fact that, as we approach the maximum possible distance from the 
site in question, the ``correction factors'' can vary substantia%
lly from unity. Of greatest practical importance is the case of 
the first-derivative operator for lattices with an even number o%
f sites, because this is the situation we are usually considerin%
g for lattice field theory. In this case, the ``correction facto%
rs'' actually vanish as we approach the maximum distance. This m%
eans that, in an ``ideal'' implementation (which wouldn't care a%
bout the cost of floating-point operations), the probability of 
performing a calculation at large $n$ will fall off much faster 
than the $1/n$ of the infinite-lattice operator. By maintaining 
the probability at $1/n$, we would be computing a greater number 
of linkages than we need to: the extra linkages would come in wi%
th a weight substantially less than unity, which means that they 
are wasteful of computing power when we consider the balance bet%
ween statistical noise and the number of computations performed.%
\par The solution is to slightly change the philosophy of the co%
mputation of the probabilistic weight and the correction factor. 
Namely, we forget about the particular structure of the SLAC der%
ivative operator altogether, and consider the case of \mbox{}%
\protect\/{\protect\em any\protect\/} arbitrary probability $p$, 
where \raisebox{0ex}[1ex][0ex]{$\protect\displaystyle0<p\leq1$}. 
We now compute \raisebox{0ex}[1ex][0ex]{$\protect\displaystyle M%
\:\!\!\equiv\mbox{round}(1/p)$}, where \raisebox{0ex}[1ex][0ex]{%
$\protect\displaystyle\mbox{round}(x)$} is the closest integer t%
o $x$. In the innermost lattice loop, we efficiently generate a 
random integer between $0$ and \raisebox{0ex}[1ex][0ex]{$\protect
\displaystyle M^{\!}-1$}, as described in [\ref{cit:Costella2002%
a}]. If this random integer vanishes, we perform the linkage cal%
culation, and multiply it by a weight of \raisebox{0ex}[1ex][0ex%
]{$\protect\displaystyle\!M^{\:\!\!}p$}. This process then ensur%
es that we are \mbox{}\protect\/{\protect\em always\protect\/} p%
erforming a linkage that has a weight that is close to unity, an%
d hence is maximally efficient. (The greatest deviation from uni%
t weight occurs for high probabilities: For \raisebox{0ex}[1ex][%
0ex]{$\protect\displaystyle p>2/3$}, we find that \raisebox{0ex}%
[1ex][0ex]{$\protect\displaystyle M^{\!}=1$} and hence the linka%
ge is \mbox{}\protect\/{\protect\em always\protect\/} performed, 
with a weight that is less than unity, but no less than $2/3$. F%
or \raisebox{0ex}[1ex][0ex]{$\protect\displaystyle2/5<p\leq2/3$}%
, we find that \raisebox{0ex}[1ex][0ex]{$\protect\displaystyle M%
^{\!}=2$}, so that the calculation is performed half the time, w%
ith a weight that can lie in the range \raisebox{0ex}[1ex][0ex]{%
$\protect\displaystyle4/5<M^{\:\!\!}p\leq4/3$}. For \raisebox{0e%
x}[1ex][0ex]{$\protect\displaystyle p\leq2/5$} the weight will c%
learly not vary from unity by more than 20\%.)\par Simulations o%
f the stochastic version of the SLAC derivative operator, using 
this approach, have successfully demonstrated the expected behav%
iour. The Fourier transform of the operator has a mean value of 
precisely $ip$, with statistical noise that is distributed almos%
t completely uniformly between all of the momentum-space compone%
nts (except \raisebox{0ex}[1ex][0ex]{$\protect\displaystyle p=0$%
}, which vanishes by symmetry for odd $N$, and has a slightly la%
rger value than the other momentum components for even $N$).\par
As a final note, the modified probabilistic algorithm described 
above allows us to ``dial the stochasticity''\raisebox{0ex}[1ex]%
[0ex]{$\protect\displaystyle\:\!\!$} of the given operator, if d%
esired. To do this, we modify the computation of $M$ to \raisebox
{0ex}[1ex][0ex]{$\protect\displaystyle M\:\!\!\equiv\mbox{round}%
(\eta/p)$}. If the resulting $M$ vanishes, then we set it equal 
to unity. If we set the dial to \raisebox{0ex}[1ex][0ex]{$%
\protect\displaystyle\eta=1$}, we obtain the probabilistic algor%
ithm described above. If we set the dial to \raisebox{0ex}[1ex][%
0ex]{$\protect\displaystyle\eta=0$}, then we will always obtain 
\raisebox{0ex}[1ex][0ex]{$\protect\displaystyle M^{\!}=1$} (pipp%
ed up from \raisebox{0ex}[1ex][0ex]{$\protect\displaystyle M^{\!%
}=0$}), and so we will always perform the linkage, which means t%
hat we have removed completely the stochastic nature of the oper%
ator. For \raisebox{0ex}[1ex][0ex]{$\protect\displaystyle0<\eta<%
1$} we have an intermediate situation, in which the statistical 
noise is less than for \raisebox{0ex}[1ex][0ex]{$\protect
\displaystyle\eta=1$}, but not zero. Thus the parameter $\eta$ i%
s effectively the ``stochasticity''\mbox{$\!$}, a number that we 
can dial. We can even set \raisebox{0ex}[1ex][0ex]{$\protect
\displaystyle\eta>1$}, which will perform linkages even less oft%
en, at the expense of correspondingly increasing the weights abo%
ve unity, so that the statistical noise is increased.\par\vspace
{1.5\baselineskip}\par{\centering\bf References\\*[0.5%
\baselineskip]}{\protect\mbox{}}\vspace{-\baselineskip}\vspace{-%
2ex}\settowidth\CGDnum{[\ref{citlast}]}\setlength{\CGDtext}{%
\textwidth}\addtolength{\CGDtext}{-\CGDnum}\begin{list}{Error!}{%
\setlength{\labelwidth}{\CGDnum}\setlength{\labelsep}{0.75ex}%
\setlength{\leftmargin}{\labelwidth}\addtolength{\leftmargin}{%
\labelsep}\setlength{\rightmargin}{0ex}\setlength{\itemsep}{0ex}%
\setlength{\parsep}{0ex}}\protect\frenchspacing\setcounter{CBtnc%
}{1}\addtocounter{CBcit}{1}\item[\hfill{[}\arabic{CBcit}{]}]%
\renewcommand\theCscr{\arabic{CBcit}}\protect\refstepcounter{Csc%
r}\protect\label{cit:Costella2002a}J.~P.~Costella, \renewcommand
\theCscr{Costella}\protect\refstepcounter{Cscr}\protect\label{au%
:Costella2002a}\renewcommand\theCscr{2002a}\protect
\refstepcounter{Cscr}\protect\label{yr:Costella2002a}{}%
\verb+hep+%
\verb+-lat/0207008+.\addtocounter{CBcit}{1}\item[\hfill{[}\arabic
{CBcit}{]}]\renewcommand\theCscr{\arabic{CBcit}}\protect
\refstepcounter{Cscr}\protect\label{cit:Costella2002b}J.~P.~Cost%
ella, \renewcommand\theCscr{Costella}\protect\refstepcounter{Csc%
r}\protect\label{au:Costella2002b}\renewcommand\theCscr{2002b}%
\protect\refstepcounter{Cscr}\protect\label{yr:Costella2002b}{}%
\verb+hep+%
\verb+-lat/0207015+.\addtocounter{CBcit}{1}\item[\hfill{[}\arabic
{CBcit}{]}]\renewcommand\theCscr{\arabic{CBcit}}\protect
\refstepcounter{Cscr}\protect\label{cit:Costella2004}J.~P.~Coste%
lla, \renewcommand\theCscr{Costella}\protect\refstepcounter{Cscr%
}\protect\label{au:Costella2004}\renewcommand\theCscr{2004}%
\protect\refstepcounter{Cscr}\protect\label{yr:Costella2004}{}%
\verb+hep+%
\verb+-lat/0404009+.\renewcommand\theCscr{\arabic{CBcit}}\protect
\refstepcounter{Cscr}\protect\label{citlast}\settowidth\Cscr{~[%
\ref{cit:Costella2004}]}\end{list}\par\end{document}